%% file: Main.tex
\documentclass[lettersize,journal]{IEEEtran}

\usepackage[utf8]{inputenc}

\input paperdefs.tex

\begin{document}
\acrodef{DER}{Distributed Energy Resource}
\acrodef{IBR}{Inverter-Based Resource}
\acrodef{LV}{Low-Voltage}
\acrodef{MV}{Medium-Voltage}
\acrodef{TSO}{Transmission System Operator}
\acrodef{DSO}{Distribution System Operator}
\acrodef{DN}{Distribution Network}
\acrodef{TN}{Transmission Network}
\acrodef{LC}{Local-Controller}
\acrodef{VDER}{Virtual \ac{DER}}
\acrodef{CA}{Control Area}
\acrodef{RoU}{Rate-of-Update}
\acrodef{GNE}{Generalized Nash Equilibrium}
\acrodef{LPF}{Low-pass Filter}
\acrodef{LP}{Linear Programming}
\acrodef{SOCP}{Second-Order Cone Programming}
\acrodef{DOPF}{Distribution Optimal Power Flow}
\acrodef{BFS-DOPF}{Backward/Forward Sweep-\ac{DOPF}}
\acrodef{MPC}{Model Predictive Control}
\acrodef{PNM}{Projected Newton Method}
\acrodef{LOOD}{Lyapunov Optimization-Based Online Distributed}
\acrodef{ADN}{Active Distribution Network}
\acrodef{PGD}{Projected Gradient Decent}
\acrodef{IMPC}{Integral \ac{MPC}}
\acrodef{EV}{Electric Vehicle}
\acrodef{PV}{Photovoltaic}
\acrodef{ADMM}{Alternating Direction Method of Multipliers}
\acrodef{DMS}{Distribution Management System}
\acrodef{DERMS}{\ac{DER} Management System}

\title{A Multi-Area Architecture for Real-Time Feedback-Based Optimization of Distribution Grids}


\author{Ilyas Farhat,~\IEEEmembership{Student Member,~IEEE,} Etinosa Ekomwenrenren,~\IEEEmembership{Student Member,~IEEE,} John W. Simpson-Porco~\IEEEmembership{Senior Member,~IEEE} Evangelos Farantatos,~\IEEEmembership{Senior Member,~IEEE,} Mahendra Patel,~\IEEEmembership{Fellow,~IEEE,} Aboutaleb Haddadi,~\IEEEmembership{Senior Member,~IEEE,} 
\thanks{Ilyas Farhat and Etinosa Ekomwenrenren are with the Department of Electrical and Computer Engineering, University of Waterloo, Waterloo, ON N3L 3G1, Canada (e-mail: ilyas.farhat@uwaterloo.ca;
etinosa.ekomwenrenren@uwaterloo.ca).}

\thanks{Evangelos Farantatos, Mahendra Patel, and Aboutaleb Haddadi are with the Electric Power Research Institute, Palo Alto, CA 94304 USA (e-mail: efarantatos@epri.com; mpatel@epri.com; ahaddadi@epri.com).}
\thanks{John W. Simpson-Porco is with the Department of Electrical and Computer Engineering, University of Toronto, Toronto, ON M5S 3G4, Canada (e-mail:jwsimpson@ece.utoronto.ca).}}

\markboth{Submitted for Publication, January~2024.}%
{Shell \MakeLowercase{\textit{et al.}}: A Sample Article Using IEEEtran.cls for IEEE Journals}


\maketitle

\begin{abstract}
A challenge in transmission-distribution coordination is how to quickly and reliably coordinate \acfp{DER} across large multi-stakeholder \acfp{DN} to support the \acf{TN}, while ensuring operational constraints continue to be met within the \ac{DN}. 
Here we propose a hierarchical feedback-based control architecture for coordination of \acp{DER} in \acp{DN}, enabling the \ac{DN} to quickly respond to power set-point requests from the \ac{TSO} while maintaining local \ac{DN} constraints. 
Our scheme allows for multiple independently-managed areas within the \ac{DN} to optimize their local resources while coordinating to support the \ac{TN}, and while maintaining data privacy; the only required inter-area communication is between physically adjacent areas within the \ac{DN} control hierarchy. 
We conduct a rigorous stability analysis, establishing intuitive conditions for closed-loop stability, and provide detailed tuning recommendations. 
The proposal is validated via case studies on multiple feeders, including IEEE-123 and IEEE-8500, using a custom MATLAB\textsuperscript{\textregistered}-based application which integrates with OpenDSS\textsuperscript{\copyright}. The simulation results show that the proposed structure is highly scalable and can quickly coordinate \acp{DER} in response to \ac{TSO} commands, while responding to local disturbances within the \ac{DN} and maintaining \ac{DN} operational limits.
\end{abstract}

\begin{IEEEkeywords}
Distribution network resources, decentralized control, smart grid, next generation control, distributed energy resources control, multi-area control, DERs coordination in distribution networks
%
\end{IEEEkeywords}

\section{Introduction}


\IEEEPARstart{T}he grid is currently transitioning from being primarily powered by fossil-fuel generators with substantial rotating inertia to a more environmentally-friendly grid predominantly powered by renewable inverter-based resources (IBRs) with decoupled rotational dynamics. This evolution introduces added complexity to the operational strategies essential for the reliable management of the bulk grid. Increased IBR penetration leads to issues such as large and more frequent frequency and voltage excursions; for example, research by the California ISO (CAISO) underscores that the uncertainties and variability from these IBRs will require faster and more flexible regulation services to maintain system reliability \cite{abdul2012enhanced}. 

Coordinated control of \acfp{DER} \textemdash{} including battery/thermal storage, distributed generation, flexible load, and electric vehicles \textemdash{} is one of the most promising solutions to address these challenges. Indeed, DERs (or aggregations thereof) offer unique advantages for fast regulation services, including: low capital costs, integration within existing load centers, and (depending on the resource) fast response to commands \cite{FERC2020}. As DERs are primarily distribution-connected \cite{dubey2023distribution}, and are owned and operated by a variety of stakeholders, harnessing their full collective flexibility for emergency and ancillary services to the bulk grid requires further advances in \ac{TN}--\ac{DN} coordination \cite{FERC2020, dubey2023distribution}. In particular, there is a need for a hierarchical framework that enables safe and fast coordination of DERs and DER aggregates behind the substation, while respecting operational boundaries and stakeholder privacy. 
\smallskip
\paragraph*{Literature Review} For the goal of providing fast ancillary services to the bulk grid, a candidate control architecture for large-scale real-time DER coordination should meet the following performance and practicality requirements: (i) the framework must coordinate resources optimally on a time-scale of seconds or faster\footnote{This requirement is strongly motivated by the authors' recent work on fast frequency and voltage regulation using transmission-connected inverter-based resources \cite{ekomwenrenren2021hierarchical,ZT-EE-JWSP-EF-MP-HH:20l,ekomwenrenren2022integrated}; fast DER coordination will be required for any \ac{TN}-\ac{DN} coordination scheme to emulate the control performance of an equivalent transmission-connected resource.}, (ii) the design should not depend on detailed system and component models, (iii) the framework should preferentially use local models, measurements, and communication, with flexibility on the amount of measurement feedback available\footnote{Critically, given that current data acquisition and communication infrastructures at the distribution level are under development, and may be cost-prohibitive to deploy extensively \cite{FERC2020}.}, and (iv) operational boundaries behind the substation should be maintained, as should the privacy of stakeholders, by minimizing the need for sharing of models and data. We next provide an overview of some recently proposed coordination methodologies, evaluating them against these criteria.

In the pursuit of efficient DER coordination for providing fast ancillary services, centralized optimization techniques have been a subject of extensive study. Centralized feeder control approaches have been based on \ac{LP} and \ac{SOCP} for \ac{DOPF} \cite{RRJ-AD:21}, Linearized \ac{DOPF} \cite{SK-CM-PA-GH:21}, \ac{MPC} \cite{HTN-DHC:23} and feedback-based optimization \cite{AB-ED:19}. While centralized optimization techniques can in principle achieve globally optimal coordination of heterogeneous DERs, they may struggle to satisfy the other desired criteria (i)--(iv) described above. Centralized techniques inherently require data sharing with a central controller, which raises fundamental concerns regarding data privacy, potentially violating requirements (iii) and (iv). Furthermore, especially in large distribution feeders, operating on a time scale of seconds or less becomes infeasible due to limitations in the communication infrastructure, which violates requirement (i) \cite{LB-MRJ-LS:23} (and references [1] and [2] therein).

Several recent studies \cite{HTN-DHC:23, RG-FS-MP:21, LW-AD-AHG-AKS-NS:22,XC-ED-CZ-NL:20, XC-NL:21,AI-SK:23,BF-HI-CB:21,GC-AB-RC-SZ:22,SF-GH-XZ-MC:21}, have aimed to overcome the limitations of centralized \ac{DER} coordination methods by leveraging decentralized optimization algorithms. While the specific \ac{DER} coordination problems differ in these references, we focus on the proposed coordination architectures. 
Both \cite{HTN-DHC:23} and \cite{RG-FS-MP:21} propose decentralized control structures based on centralized \ac{MPC}, which are then decentralized using \ac{ADMM}. In \cite{RG-FS-MP:21}, a power set-point tracking problem is considered wherein a pre-scheduled set-point curve is planned through a stochastic optimization problem, and the central coordinator dispatches the \acp{DER}. This however occurs over a timescale of 30s, and does not consider real-time disturbance regulation. On the other hand, \cite{HTN-DHC:23} considers the problem of coordinating \ac{EV} charging stations to minimize the costs while ensuring stable distribution grid operation. 
Similarly, \cite{LW-AD-AHG-AKS-NS:22} proposes a decentralized voltage control structure for distribution systems with \ac{PV} and \ac{EV} \acp{DER}. This structure include a dynamic network clustering based on network structure, voltage sensitivities and regulation capabilities. Additionally, each cluster is equipped with a \ac{MPC}-based voltage controller. 
In \cite{XC-ED-CZ-NL:20} and \cite{XC-NL:21}, a \ac{DER} coordination scheme is proposed through a distributed MPC framework. This scheme relies on a method for modeling and quantifying the aggregate power flexibility within the feeder. The primary goal is to maintain the privacy of \acp{DER} by only sharing information about their flexibility. The proposed control structures in \cite{HTN-DHC:23, RG-FS-MP:21, LW-AD-AHG-AKS-NS:22,XC-ED-CZ-NL:20, XC-NL:21} successfully limit data sharing with the DSO controller, but still involve centralized coordination and do not use real-time feedback to maintain \ac{DN} operating constraints.

Steps towards a true hierarchical framework were taken in \cite{AI-SK:23} and \cite{BF-HI-CB:21}, which propose two-level control architectures to coordinate low-voltage \acp{DER}. In \cite{AI-SK:23}, a primary-secondary (leader-follower) framework using decentralized \ac{IMPC} is introduced, with a focus on maintaing data privacy. This framework serves as an interface layer connecting medium-voltage and low-voltage networks for coordinating low-voltage \acp{DER}. In contrast, \cite{BF-HI-CB:21} proposes a hierarchical distributed structure that centrally updates the set-points for low-voltage \acp{DER}. In both architectures, data-sharing between the layers/controllers is minimized, but both lack the flexibility to partition the feeder into multiple areas, which is essential for preserving stakeholder data privacy and operation when multiple operators are present in large low-voltage networks. 

In \cite{GC-AB-RC-SZ:22} a decentralized feedback-based method is proposed for solving \ac{DN} OPF problems with prosumers. This approach leverages local voltage measurements within a gradient descent method, while minimizing communication with a central entity. In \cite{SF-GH-XZ-MC:21}, a distributed control structure is introduced, utilizing a \ac{LOOD} algorithmic framework for \acp{ADN}. The framework effectively controls numerous \acp{DER} through the \ac{ADN} operator, acting as a central coordinator. The \ac{ADN} operator receives set-point reference signals from the \ac{TSO} and tracks them by sending incentive signals to the \acp{DER}. Each \ac{DER} relies on local information and measurements to update their power set-points. Similarly, an online feedback-based algorithm is proposed in \cite{AB-ED:19} to coordinate \acp{DER} within \ac{DN} while ensuring  circuit constraints are maintained. The algorithm leverages online projected-gradient methods to track the solution of a time-varying optimization problem implemented by a centralized controller to track requested set-points from the \ac{TSO}. In contrast, \cite{SN-YCC-LW:20} proposes a centralized control scheme, where the controller estimates \acp{DER}' sensitives in real-time using network measurements. Additionally, the controller updates \acp{DER} set-points to control and track voltages and powers at network buses, ensuring they remain within limits. However, the centralized nature in \cite{AB-ED:19,GC-AB-RC-SZ:22, SF-GH-XZ-MC:21, SN-YCC-LW:20} architectures violates requirements (iii) and (iv).

Following the overview of various studies that explored centralized, and decentralized structures, we pose our proposed structure as a candidate that leverages recent advances in control theory, specifically so-called \emph{feedback-based optimization} \cite{0155_Colombino,colombino2019online,AB-ED:19, LSPL-JWSP-EM:18l, 0154_ORTMANN2020106782, EDA-SVD-GBG:15, GB-JC-JIP-ED:21, ZT-EE-JWSP-EF-MP-HH:20l,AB-JC-YC-JW:23} to optimally coordinate DERs. The decentralized control structure proposed, along with feedback-based optimization, serves as control structure that promises to fulfill the design requirements (i)-(iv) for distribution networks.

\smallskip


\paragraph*{Statement of Contributions}  We propose a \emph{hierarchical and multi-area} control architecture to quickly coordinate distribution level DERs and DER aggregates. At a macro-level, the control scheme receives active and reactive power set-point commands from the \ac{TSO}, and optimally redispatches \acp{DER} within the \ac{DN} to track the set-point, while leveraging real-time feedback from the grid to ensure satisfaction of line current and bus voltage constraints within the \ac{DN}. The use of measurement feedback renders the scheme inherently robust against inaccuracy in the \ac{DN} grid model used for the design.
 
The most innovative aspect of our approach compared to the literature (e.g., \cite{AB-ED:19}) is that it provides a highly scalable multi-area control architecture. At a micro level, the \ac{DN} is divided into a hierarchy of local control areas. Local controllers manage all \acp{DER} and \ac{DN} measurements for each area to preserve operational boundaries, and each local controller design requires only local \ac{DER} capacity information and an approximate local \ac{DN} model, preserving stakeholder privacy. Coordination between controllers is minimal, and is hierarchical to minimize long-distance communication. This architecture addresses the challenge of a multi-stakeholder system and significantly reduces overall communication requirements. 
We provide a rigorous closed-loop stability analysis, deriving intuitive and explicit analytic conditions for stability, and provide a systematic and practical tuning procedure for the design. Finally, we validate the proposed scheme via case studies on several feeders, including the IEEE-123 and IEEE-8500 feeders, and describe several modifications which have been found to improve performance in implementation. Simulation results indicate that our scheme can provide fast and highly scalable coordination of \acp{DER} in large multi-stakeholder \acp{DN}.

\smallskip

\paragraph*{Paper Organization} Section \ref{sec: Multi-level Control Structure} outlines the multi-level control architecture, presenting the distribution network model and other preliminaries. Section \ref{sec: Problem Formulation} presents the local optimization problem of each \acf{LC} and the proposed controller. A closed-loop stability analysis is presented in Section \ref{Sec:Theory}, along with comprehensive guidelines for tuning the controllers' parameters. Case studies on three feeders are presented in Section \ref{sec: Simulations}, with conclusions and future directions described in Section \ref{sec: Conclusions}.

\section{Overview of Proposed Hierarchical Control Structure}\label{sec: Multi-level Control Structure}

\begin{figure}[ht!]
    \centering
    \includegraphics[width=\columnwidth]{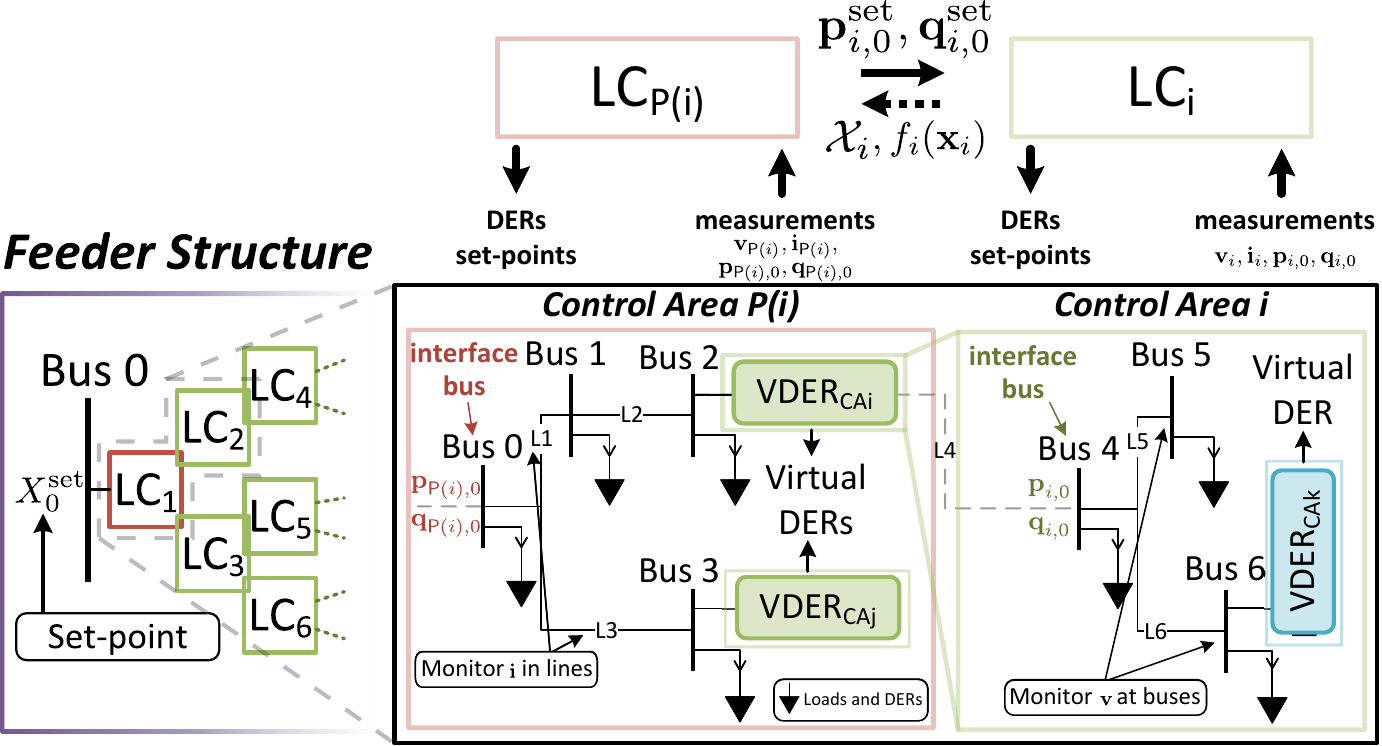}
    \caption{Single-feeder \ac{DN} internal structure. Zoomed-in areas illustrate the interaction between parent and child areas within a feeder. Red and green boxes illustrate the visibility of each \ac{LC} over the local infrastructure and resources, including local grid measurements ($\rm \mathbf{v,i,p,q}$) and local \acp{DER}.}
    \label{Fig:Hierarchical and Parent-Child}
\end{figure}

Our focus is on the control of single-feeder\footnote{If multiple feeders are connected to the \ac{TN}-\ac{DN} interface, the TSO power request can be allocated across the feeders, and our subsequent control design can then be applied independently to each feeder.} \ac{DN} which has been partitioned into a hierarchy of areas, as shown in Figure \ref{Fig:Hierarchical and Parent-Child}. The high-level control objective is to leverage controllable DERs within the \ac{DN} such that (i) an overall power set-point $X_0^{\rm set} = (\mathrm{p}_{0}^{\mathrm{set}}, \mathrm{q}_{0}^{\mathrm{set}}) \in \real^2$ provided by the \ac{TSO} is tracked at the \ac{TN}-\ac{DN} interface bus, (ii) operational constraints are maintained throughout the DN, and (iii) DERs are used efficiently. Additionally \textemdash{} and as the most distinguishing feature of this work \textemdash{} the design is subject to the constraint that information and management boundaries between different areas of the DN must be respected. This leads to a hierarchical area-based control architecture to be described next.

\subsection{Feeder Architecture and \acfp{LC}}
\label{Sec:FeederArchitecture}

%
%
We consider the feeder as being partitioned into $N$ \acp{CA}. These \acp{CA} may represent contractual arrangements for the management of \ac{DER} resources via aggregators, or may be defined based on other operational criteria such as communication capabilities or having uniform division of control resources within the areas \cite{0165_Han2014,0166_Utkarsh2022}. Each \ac{CA} will be controlled by a controller, which will have some visibility over the electrical infrastructure within that area and will redispatch local DERs. Importantly, the controller will \emph{not} have visibility of neighboring \acp{CA}, and hence the control structure will be decentralized.

The arrangement of \acp{CA} and the communication between the associated \acp{LC} will be hierarchical. This ensures that (i) any required communication is local, which minimizes communication delays, (ii) the control architecture is scalable to large numbers of control areas, and (iii) the majority of local measurements and area data are not shared to maintain the privacy of stakeholders within the area. We describe the hierarchical arrangement between the \acp{CA} with a directed rooted tree graph $\mathcal{G}_{\rm CA} = (\mathcal{N}_{\rm CA},\mathcal{E}_{\rm CA})$ with nodes $\mathcal{N}_{\rm CA} = \{1,\ldots,N\}$ and edges $\mathcal{E}_{\rm CA}$; the root of the tree is the first \ac{CA} at the head of the feeder; see Figure \ref{Fig:Hierarchical and Parent-Child}. For any \ac{CA} $i \in \mathcal{N}_{\rm CA}$, we let $\mathsf{P}(i)$ denote the unique parent \ac{CA} and $\mathsf{C}(i)$ denote the set of child \acp{CA}; by convention $\mathsf{P}(1) = \emptyset$ and $\mathsf{C}(N) = \emptyset$. We let $\mathcal{A} \in \{0,1\}^{N \times N}$ denote the adjacency matrix of $\mathcal{G}_{\rm CA}$, with elements $\mathcal{A}_{ij} = 1$ if $(i,j) \in \mathcal{E}_{\rm CA}$ and 0 otherwise. 
 
Any control area $i \in \mathcal{N}_{\rm CA}$ is connected to its parent area $\mathsf{P}(i)$ through a single bus which we call the \emph{interface bus}; see Figure \ref{Fig:Hierarchical and Parent-Child}. From the perspective of the parent area $\mathsf{P}(i)$, the $i$th \ac{CA} is represented as a \acf{VDER} located at this interface bus. The \ac{VDER} is a fictitious \ac{DER} which will be ``dispatched'' by the parent area \ac{LC} by providing a power set-point $X_{i,0}^{\rm set} = (\mathrm{p}_{i,0}^{\rm set}, \mathrm{q}_{i,0}^{\rm set})$ to the \ac{LC} of area $i$.\footnote{The set-point for \ac{CA} $1$ at the head of the feeder is provided by the \ac{TSO}.} The control cost $f_i$ and capacity limits $\boldsymbol{\mathcal{X}}_i$ of this \ac{VDER} (i.e., of the child area) are assumed known by the LC of the parent area, and may be updated by exception through communication from the child to the parent area; further discussion on capacity limits and cost functions is deferred to Section \ref{sec: LC Optimization Problem}. Within the \ac{LC}, the objective is then to track this provided set-point at the interface bus by redispatching local DERs, while maintaining local constraints on voltages and currents. This tracking control will be accomplished through a combination of local optimization and feedback, to be described in Section \ref{sec: LC Optimization Problem}.

\paragraph*{\acs{DMS} and \acs{DERMS} Coordination within Our Hierarchical Control Framework} In the evolving landscape of power distribution systems, the proposed hierarchical structure seamlessly aligns with the distinct functionalities of \acf{DMS} and \acf{DERMS}. Notably, the coordination between these modules is vital for effective \ac{DER} management. In essence, \ac{DERMS} assumes the critical role of overseeing and managing \acp{DER} under its jurisdiction, responding to power requests from the \ac{DMS}. The \ac{DMS} module, associated with the \ac{CA}, accesses crucial information including circuit model, measurements and \acp{DER} locations. However, the intricate details of \acp{DER} \textemdash{} internal structures, operating limits, and specific models \textemdash{} are intentionally withheld from the \ac{DMS}. On the flip side, \ac{DERMS} emerges as the communication hub directly interfacing with \acp{DER}, equipped with comprehensive insights into their constraints, internal models and communication requirements.

This collaborative framework enables \ac{DMS} to provide \acp{DERMS} with critical data, including \acp{DER} sensitivities (to be described in Section \ref{sec: Bernstein Model}) and updated dual variables corresponding to power tracking and circuit constraints (see Algorithm \ref{Alg:1} for details). Meanwhile, \ac{DERMS} equipped with updated duals and \acp{DER} sensitivities recieved from \ac{DMS} (without sharing circuit model or measurements), issues set-points to \acp{DER} in response. Within the proposed structure, each \ac{CA} is potentially managed by a distinct stakeholder and equipped with their own \ac{DMS} and \acp{DERMS}\footnote{At any \ac{CA}, their might be \ac{DER} service providers that are independent from the stakeholder, and hence would have their own \ac{DERMS} module that can be integrated with the \ac{CA}'s \ac{DMS}.}. For more details about \ac{DERMS} and \ac{DMS} functionalities and their intricate coordination, refer to \cite{BS-EPRI:18,BD-EPRI:13,AM-TH-JW-RS-NK-XL-JR-AP-SV:17}.

\subsection{Model of a Single Control Area}
\label{Sec:SingleControlArea}

We now describe the model for the distribution system and the DER components contained within a given control area. 

\smallskip

\subsubsection{Controllable DERs} We denote by $\mathcal{D}_i$ the set of controllable \acp{DER} (including any \acp{VDER}) within the \ac{CA} $i \in \mathcal{N}_{\rm CA}$. The \acp{DER} are assumed to have local controllers which allow them to quickly track set-points provided by the \ac{LC}. For any \ac{DER} $j\in\mathcal{D}_i$, we let $x_j=(p_j,q_j) \in \real^2$ denote its overall active and reactive power set-points, which are constrained to be within capacity limits specified by the closed, non-empty convex set $\mathcal{X}_j\subset \real^2$. We let $\mbf{x}_i = \mathrm{col}(x_1,\dots,x_{|\mathcal{D}_i|}) \in \real^{2|\mathcal{D}_i|}$ denote the stacked vector of all \ac{DER} power set-points for the $i$th CA, and is subject to the overall limits $\mbf{x}_i\in\bs{\mc{X}}_i \define \mathcal{X}_1\times\cdots\times\mathcal{X}_{|\mathcal{D}_i|}$.

\subsubsection{Distribution Network Model}\label{sec: Bernstein Model}
Let $\overline{\mathcal{N}}_{i} = \{0\} \cup \mathcal{N}_{i}$ with $\mathcal{N}_{i} := \{1,2,\dots,N_i\}$ denote the set of buses in the $i$th \ac{CA}, where the interface bus of the \ac{CA} is given the node ``0''; each bus is potentially multi-phase, with up to three phases. Not necessarily all buses and lines within the \ac{CA} will be monitored for control purposes; we let $\mathcal{M}^{\mathbf{v}}_{i} \subseteq \mathcal{N}_{i}$ denote the set of buses where phase-to-ground voltage magnitude measurements are available, with the understanding that all phases will be monitored if the bus is multi-phase, and let $\mathbf{v}_i$ be the vector of measured voltages. Similarly, we let $\mathbf{i}_{i}$ denote the vector of measured line current magnitudes for a subset of monitored distribution lines $\mathcal{M}^{\mathbf{i}}_{i} \subseteq \overline{\mathcal{N}}_i \times \overline{\mathcal{N}}_i$. Finally, we let $\mathbf{p}_{i,0}, \mathbf{q}_{i,0}\in\real^{m_i}$ denote the net active and reactive power injections at the interface bus, where $m_i$ is the number of phases at the interface bus. 

We adopt the distribution network model from \cite{0057_FPL_Bernstein_followup,0059_FPL_Bernstein_Parent}, which is capable of modelling both radial and meshed unbalanced networks, as well as wye and delta connections for loads and \acp{DER}. The model provides linearized equations relating the \acp{DER}' powers to the \ac{CA}'s voltages and currents measurements, and to the interface bus power injections. In the current work, we have adjusted these equations to combine both wye and delta-connected \acp{DER}, leading to the model
\begin{subequations}\label{eq:DN_Equations}
\begin{align}
    \label{eq:DN_Equations_1}
    \mathbf{v}_i (\mbf{x}_i) &= \mathbf{A}_i \mbf{x}_i + \mathbf{a}_i\\
    \label{eq:DN_Equations_2}
    \mathbf{i}_i (\mbf{x}_i) &= \mathbf{B}_i \mbf{x}_i + \mathbf{b}_i\\
    \label{eq:DN_Equations_3}
    \mathbf{p}_{i,0} (\mbf{x}_i) &= \mathbf{M}_i \mbf{x}_i + \mathbf{m}_i\\
    \label{eq:DN_Equations_4}
    \mathbf{q}_{i,0} (\mbf{x}_i) &= \mathbf{H}_i \mbf{x}_i + \mathbf{h}_i
\end{align}
\end{subequations}
where $\mathbf{A}_i$, $\mathbf{B}_i$, $\mathbf{M}_i$ and $\mathbf{H}_i$ are constant matrices which can be computed from the operating point, the admittance matrix, and the specification of DER phase connections, and $\mathbf{a}_i$, $\mathbf{b}_i$, $\mathbf{m}_i$ and $\mathbf{h}_i$ are constant vectors. For more details about the linearization, please refer to the supplement \cite{FarhatSupplOnlineMat}. Note that the matrices $\mathbf{A}_i$, $\mathbf{B}_i$, $\mathbf{M}_i$ and $\mathbf{H}_i$ can be interpreted as sensitivity matrices; for instance, $\mathbf{A}_i$ captures the sensitivity between \ac{DER} set-points and voltage magnitudes at the measurement points. 


\section{Hierarchical Feedback-Based Optimization of Distribution Feeders}\label{sec: Problem Formulation}

\subsection{\ac{LC} Optimization Problem}
\label{sec: LC Optimization Problem}

The control objectives of the \ac{LC} for the $i$th \ac{CA} will be formulated by specifying an optimization problem. The problem aims to re-dispatch \acp{DER} within their limits to track provided power set-points $X_{i,0}^{\rm set} = (\mathrm{p}_{i,0}^{\rm set}, \mathrm{q}_{i,0}^{\rm set})$ at the interface bus, while maintaining measured voltages $\mbf{v}_i$ and currents $\mbf{i}_i$ within constraints, and by efficiently using \ac{DER} resources. As described in Section \ref{Sec:FeederArchitecture}, the set-points $\mathrm{p}_{i,0}^{\rm set}$ and $\mathrm{q}_{i,0}^{\rm set}$ to be tracked will be computed as VDER set-points by the parent area, which we express as
\begin{equation}\label{Eq:ParentSetPoints}
\mathrm{p}_{i,0}^{\rm set}(\mathbf{x}_{\mathsf{P}(i)}) = T^{\rm p}_{i}\mathbf{x}_{\mathsf{P}(i)}, \quad \mathrm{q}_{i,0}^{\rm set}(\mathbf{x}_{\mathsf{P}(i)}) = T^{\rm q}_{i}\mathbf{x}_{\mathsf{P}(i)}
\end{equation}
for appropriate matrices $T^{\rm p}_{i}$ and $T^{\rm q}_{i}$. Mathematically, we can now express the optimization problem for the $i$th CA as
\begin{subequations}\label{eq:LC-Problem}
\begin{align}\label{Eq:LCCost}
        \minimize_{\mbf{x}_i \in \bs{\mathcal{X}}_i}  \,\,f_i(\mbf{x}_i) \define \sum_{j \in \mathcal{D}_i} f_{ij}(x_j)
\end{align}
subject to
\begin{align}
    \label{eq:LC-Constraint3}
    & s_i \left( | \vones^{\T} \mathbf{p}_{i,{0}} (\mbf{x}_i) - \mathrm{p}_{i,0}^{\rm set}(\mathbf{x}_{\mathsf{P}(i)}) | \right) \leq E_{i_{\rm p}}\\
    \label{eq:LC-Constraint5}
    & s_i \left( | \vones^{\T} \mathbf{q}_{i,{0}} (\mbf{x}_i) - \mathrm{q}_{i,0}^{\rm set}(\mathbf{x}_{\mathsf{P}(i)}) | \right) \leq E_{i_{\rm q}}\\
    \label{eq:LC-Constraint6}
    &\mathbf{v}_i (\mbf{x}_i) \leq \overline{\mathbf{v}}_i\\
    \label{eq:LC-Constraint7}
    &\underline{\mathbf{v}}_i \leq \mathbf{v}_i (\mbf{x}_i)\\
    \label{eq:LC-Constraint8}
    &\mathbf{i}_i(\mbf{x}_i)\leq \overline{\mathbf{i}}_i
\end{align}
\end{subequations}

The linear inequalities \eqref{eq:LC-Constraint3}--\eqref{eq:LC-Constraint5} enforce total active and reactive power tracking at the interface bus of the provided set-points $\mathrm{p}_{i,0}^{\rm set}$ and $\mathrm{q}_{i,0}^{\rm set}$, within specified tolerances $E_{i_{\rm p}}, E_{i_{\rm q}} > 0$. We emphasize that from the perspective of the optimization \eqref{eq:LC-Problem} for the $i$th \ac{CA}, the parent variable $\mathbf{x}_{\mathsf{P}(i)}$ is \emph{fixed}. The fixed binary variable $s_i\in\{0,1\}$ can be used to enable or disable this tracking feature. The inequalities \eqref{eq:LC-Constraint6}, \eqref{eq:LC-Constraint7} enforce upper and lower limits $\overline{\mathbf{v}}_i$ and $\underline{\mathbf{v}}_i$ on the voltage magnitudes at the measurement points, with \eqref{eq:LC-Constraint8} limiting the current magnitude along monitored lines below $\overline{\mathbf{i}}_i$. 

\paragraph*{DER Costs and Constraints} The objective function \eqref{Eq:LCCost} is a separable cost over each DER and VDER which penalizes its use for control purposes. Our only assumption will be that $f_i$ in \eqref{eq:LC-Problem} is continuously differentiable and strongly convex; we let $m_i > 0$ denote the strong convexity parameter, and by convention, the units of the cost will be $\text{W}^2$. For example, in our case studies we will use quadratic costs of the form
\begin{equation}\label{eq:DER Cost function}
    \begin{aligned}
        f_{ij}(x_j) &= x_j^\T C^{\prime\prime}_j x_j + x_j^{\T} C^{\prime}_j ,
        %
\end{aligned}
\end{equation}
where $C^{\prime\prime}_{j} \succ 0$ is a diagonal $2 \times 2$ matrix and $C^{\prime}_j \in \real^2$. These coefficients can be set to reflect the preference of using different types of \acp{DER}; larger costs will lead to lower control usage of a given DER. This can be crucial when different \acp{DER} with different characteristics (e.g., capacities or speed dynamics) are controlled together. The power capacity constraint set $\mathcal{X}_j$ for \ac{DER} $j \in \mathcal{D}_i$ may simply be the box constraint $\mathcal{X}_j = [\underline{x}_j,\overline{x}_j]$ specifying independent active and reactive power limits, or may instead encode more complex apparent power constraints via intersections of half-planes and semi-circular regions.

%

\paragraph*{\ac{VDER} Costs and Constraints} As the \acp{VDER} of the $i$th \ac{CA} provides an aggregated representation of all \acp{DER} and \acp{VDER} in the child areas $\mathsf{C}(i)$, assigning appropriate costs to \acp{VDER} is critical to ensure resources are appropriately leveraged throughout the \ac{DN}. To ensure downstream \acp{DER} provide control action, a \ac{VDER} should have lower cost compared to other \acp{DER} available within the same \ac{CA}. If $j \in \mathcal{D}_i$ is the \ac{VDER} corresponding to to child area $k \in \mathsf{C}(i)$, then a choice that roughly mimics the allocation obtained from a global centralized optimization is
\begin{equation}\label{eq:Cost of VDER}
f_{ij}(x_j) = \left(\sum_{\ell \in \mathcal{D}_k}\nolimits f_{k\ell}^*\right)^*(x_j),
\end{equation}
where denotes ${}^*$ convex conjugate. For instance, with quadratic costs \eqref{eq:DER Cost function}, \eqref{eq:Cost of VDER} evaluates to 
\[
\begin{aligned}
f_{ij}(x_j) &=  (x_j+\zeta_{k})^{\top}\left(\sum_{l\in\mathcal{D}_k}\nolimits(C_{\ell}^{\prime\prime})^{-1}\right)^{-1}(x_j+\zeta_{k})
\end{aligned}
\]
where $\zeta_k = 2\sum_{\ell\in\mathcal{D}_k}(C_{l}^{\prime\prime})^{-1}C_{\ell}^{\prime}$. 
One can interpret this formula as akin to an equivalent impedance from a parallel combination of impedances, as all downstream DERs would be used in parallel in a centralized dispatch. While the proposed \ac{VDER} cost in \eqref{eq:Cost of VDER} is developed to emulate centralized controller implementation, it is worth noting that \ac{VDER} costs can be configured in various ways. In the context of a competitive market, where multiple service providers engage in competition, setting \ac{VDER} costs can be structured to adhere to a cost-effective plan, allowing different stakeholders to compete while delivering services to the grid.






\paragraph*{Offline vs. Online Optimization} The optimization problem \eqref{eq:LC-Problem} is convex and could in principle be directly solved. However, the constants $\mathbf{a}_i, \mathbf{b}_i, \mathbf{m}_i, \mathbf{h}_i$ in the distribution system model \eqref{eq:DN_Equations}  depend on unknown real-time loading conditions and disturbances throughout the system. Even if the best available estimates are used for these quantities in \eqref{eq:LC-Problem}, implementation of the resulting set-points in the system will likely lead to constraint violation, and the system will not actively respond as disturbances change. Instead, following \cite{AB-ED:19} and inspired also by recent advances in feedback-based optimization, we pursue an iterative approach which uses real-time measurement feedback from the system in place of this model information. 

\subsection{\ac{LC} Control Algorithm}
\label{Sec:Controller}

To introduce the \ac{LC} controller, we require the \emph{regularized Lagrangian} function $L_{i}^{\rm r}$ of the problem \eqref{eq:LC-Problem}, given by
\begin{equation}\label{eq:LowerLevelLagrangianFunction}
    \begin{split}
        L_{i}^{\rm r}&(\mathbf{x}_i,\mathbf{d}_i;\mathbf{x}_{\mathsf{P}(i)}) \coloneqq  f_i(\mathbf{x}_i)\\
        &+ \lambda_i\left(s_i\left(\vones^\T \mathbf{p}_{i,0}(\mathbf{x}_i)-\mathrm{p}_{i,0}^{\rm set}(\mathbf{x}_{\mathsf{P}(i)})\right) - E_{i_{\rm p}}\right)\\
        &+ \mu_i\left(s_i\left(\mathrm{p}_{i,0}^{\rm set}(\mathbf{x}_{\mathsf{P}(i)})-\vones^\T \mathbf{p}_{i,0}(\mbf{x}_i)\right) - E_{i_{\rm p}}\right)\\
        &+ \eta_i\left(s_i\left(\vones^\T \mathbf{q}_{i,0}(\mbf{x}_i)-\mathrm{q}_{i,0}^{\rm set}(\mathbf{x}_{\mathsf{P}(i)})\right) - E_{i_{\rm q}}\right)\\
        &+ \psi_i\left(s_i\left(\mathrm{q}_{i,0}^{\rm set}(\mathbf{x}_{\mathsf{P}(i)})-\vones^\T \mathbf{q}_0(\mbf{x}_i)\right) - E_{i_{\rm q}}\right)\\
        &+ \boldsymbol{\gamma}_i^{\T}\left(\mathbf{v}_i (\mbf{x}_i) - \overline{\mathbf{v}}_i\right) +  \boldsymbol{\nu}_i^{\T}\left(\underline{\mathbf{v}}_i - \mathbf{v}_i (\mbf{x}_i)\right)\\
        &+ \boldsymbol{\zeta}_i^{\T}\left(\mathbf{i}_i(\mbf{x}_i) - \overline{\mathbf{i}}\right)\\
        &+ \tfrac{r^{\rm p}_i}{2}\|\mathbf{x}_i\|_2^2 - \tfrac{1}{2} \mathbf{d}_i^{\T}\mathbf{R}^{\rm d}_{i}\mathbf{d}_i
    \end{split}
\end{equation}
where $\mathbf{d}_i = \mathrm{col}(\lambda_i,\mu_i,\eta_i,\psi_i,\boldsymbol{\gamma}_i,\boldsymbol{\nu}_i,\boldsymbol{\zeta}_i)$ is the vector of \emph{dual variables}, $r^{\rm p}_{i} \geq 0$ is the primal regularization parameter, and 
\begin{equation}\label{Eq:Rdi}
\mathbf{R}^{\rm d}_{i} = \mathrm{blkdiag}(r_{\lambda_i},r_{\mu_i},r_{\eta_i},r_{\psi_i},r_{\gamma_i}I, r_{\nu_i}I, r_{\zeta_i}I) \succeq 0
\end{equation}
are the dual regularization parameters, with some elements being diagonal matrices of appropriate dimensions. We refer to \cite{AB-ED:19, 0156_Sumin2022} for extensive discussion on the theoretical and practical benefits of including regularization; for our purposes, these will be tuning parameters of the approach, and a systematic procedure for setting these parameters will be described in Section \ref{Sec:Tuning}. 
Our proposed \ac{CA} controller will operate online with a sampling period of $T_{\rm s} > 0$, and is outlined in Algorithm \ref{Alg:1}, where $\mathcal{P}_{\geq 0}$ denotes Euclidean projection of the argument onto the nonnegative orthant.

\begin{algorithm} 
\caption{\ac{LC} Controller for $i$th \ac{CA}}\label{Alg:1}
{
\small
\begin{algorithmic}
\State At each sampling time
\State \textbf{[Step 1]}: Receive set-points from \ac{LC} of parent area $\mathsf{P}(i)$
\[
\mathrm{p}_{i,0}^{\rm set}(\mathbf{x}_{\mathsf{P}(i)}) = T^{\rm p}_{i}\mathbf{x}_{\mathsf{P}(i)}, \quad \mathrm{q}_{i,0}^{\rm set}(\mathbf{x}_{\mathsf{P}(i)}) = T^{\rm q}_{i}\mathbf{x}_{\mathsf{P}(i)}
\]
\State \textbf{[Step 2]}: Collect local measurements $\mathbf{p}_{i,{0}}, \mathbf{q}_{i,{0}},\mathbf{v}_i, \mathbf{i}_i, $ $\bs{\mathcal{X}}_i$
\State \textbf{[Step 3]}: \ac{LC} performs the updates
{
\footnotesize
\begin{subequations}
\label{eq:LC-Alg1}
\begin{align*}
    %
    \lambda^+_i &= \mathcal{P}_{\geq 0} \left(\lambda_i + \alpha_{\lambda_i} \left(\vones^\T \mathbf{p}_{i,0} - \mathrm{p}_{i,{0}}^{\rm set} - E_{i_{\rm p}} - r_{\lambda_i} \lambda_i\right)\right)\\
    %
    \mu^+_i &= \mathcal{P}_{\geq 0} \left(\mu_i + \alpha_{\mu_i} \left(\mathrm{p}_{i,{0}}^{\rm set} - \vones^\T \mathbf{p}_{i,0} - E_{i_{\rm p}} - r_{\mu_i} \mu_i\right)\right)\\
    %
    \eta^+_i &= \mathcal{P}_{\geq 0} \left(\eta_i + \alpha_{\eta_i} \left(\vones^\T \mathbf{q}_{i,0} - \mathrm{q}_{i,{0}}^{\rm set} - E_{i_{\rm q}} - r_{\eta_i} \eta_i\right)\right)\\
    %
    \psi_i^+ &= \mathcal{P}_{\geq 0} \left(\psi_i + \alpha_{\psi_i} \left(\mathrm{q}_{i,{0}}^{\rm set} - \vones^\T \mathbf{q}_{i,0}- E_{i_{\rm q}} - r_{\psi_i} \psi_i\right)\right)\\
    %
    \boldsymbol{\gamma}^+_i &= \mathcal{P}_{\geq 0} \left(\boldsymbol{\gamma}_i + \alpha_{\gamma_i} \left(\mathbf{v}_i - \overline{\mathbf{v}}_i  - r_{\gamma_i} \boldsymbol{\gamma}_i\right)\right)\\
    %
    \boldsymbol{\nu}^+_i &= \mathcal{P}_{\geq 0} \left(\boldsymbol{\nu}_i + \alpha_{\nu_i} \left(\underline{\mathbf{v}}_i  - \mathbf{v}_i - r_{\nu_i} \boldsymbol{\nu}_i\right)\right)\\
    %
    \boldsymbol{\zeta}^+_i &= \mathcal{P}_{\geq 0} \left(\boldsymbol{\zeta}_i + \alpha_{\zeta_i} \left(\mathbf{i}_i - \overline{\mathbf{i}}_i - r_{\zeta_i} \boldsymbol{\zeta}_i\right)\right)
\end{align*}
\end{subequations}\vspace{-1em}
}
\begin{subequations}
\State \textbf{[Step 4]}: \ac{LC} updates (V)DER set-points
\begin{align*}
    \mbf{x}_i^+ &= \argmin\limits_{\mbf{x}_i\in\bs{\mathcal{X}}_i} L_{i}^{\rm r}(\mbf{x}_i,\mathbf{d}_i^+;\mathbf{x}_{\mathsf{P}(i)})
\end{align*}
\State \textbf{[Step 5]}: Transmit set-points to \acp{LC} of each child area
\begin{align*}
\mathrm{p}_{j,0}^{\rm set}(\mathbf{x}_{i}^+) = T^{\rm p}_{j}\mathbf{x}_{i}^+, \quad \mathrm{q}_{j,0}^{\rm set}(\mathbf{x}_{i}^+) = T^{\rm q}_{j}\mathbf{x}_{i}^+, \qquad j \in \mathsf{C}(i).
\end{align*}
\end{subequations}
\end{algorithmic}
}
\end{algorithm}

Algorithm \ref{Alg:1} consists of several computationally straightforward steps and involves hierarchical interaction between the \acp{CA}. The dual variables $\mathbf{d}_i = \mathrm{col}(\lambda_i,\mu_i,\eta_i,\psi_i,\boldsymbol{\gamma}_i,\boldsymbol{\nu}_i,\boldsymbol{\zeta}_i)$ are internal states of the $i$th \ac{LC}, and the dual update rules in Step 3 directly use the measurements $(\mathbf{v}_i,\mathbf{i}_i,\mathbf{p}_{i,0},\mathbf{q}_{i,0})$ and the provided set-points $\mathrm{p}_{i,0}^{\rm set}$ and $\mathrm{q}_{i,0}^{\rm set}$ in what can be interpreted as a measurement-based gradient ascent step to maximize the Lagrangian \eqref{eq:LowerLevelLagrangianFunction}. The update rules in Step 3 are parameterized by the controller gains 
\begin{equation}\label{Eq:stepSizes}
\boldsymbol{\alpha}_i \define \mathrm{blkdiag}(\alpha_{\lambda_i},\alpha_{\mu_i},\alpha_{\eta_i},\alpha_{\psi_i},\alpha_{\gamma_i}I,\alpha_{\nu_i}I,\alpha_{\zeta_i}I)
\end{equation}
which are tuning parameters; our theoretical results to follow will address the constraints on these gains, and a systematic tuning procedure will be provided in Section \ref{Sec:Tuning} for both the regularization parameters and the gains. Next, the new DER set-points $\mathbf{x}_i^+$ are computed by the \ac{LC} by solving the local optimization problem in Step 4. Note that this local optimization can be equivalently written as
\begin{equation}\label{Eq:ArgminSimplified}
{\small
\begin{aligned}
\mathbf{x}_i^+ &= \argmin\limits_{\mbf{x}_i\in\bs{\mathcal{X}}_i} f_i(\mathbf{x}_i) + \tfrac{r^{\rm p}_{i}}{2}\|\mathbf{x}_i\|_2^2+ (\lambda_i^+-\mu_i^+) s_i \vones[]^{\T}\mathbf{M}_i\mbf{x}_i\\
& + (\eta_i^+-\psi_i^+)s_i \vones[]^{\T} \mathbf{H}_i\mathbf{x}_i + (\boldsymbol{\gamma}_i^+-\boldsymbol{\nu}_i^+)^{\T}\mathbf{A}_i\mathbf{x}_i + (\boldsymbol{\zeta}_i^+)^{\T}\mathbf{B}_i\mathbf{x}_i
\end{aligned}
}
\end{equation}
and thus requires knowledge of the sensitivity matrices $(\mathbf{A}_i,\mathbf{B}_i,\mathbf{M}_i,\mathbf{H}_i)$ from \eqref{eq:DN_Equations}, but \emph{does not} require the unknown load-dependent constants $(\mathbf{a}_i,\mathbf{b}_i,\mathbf{m}_i,\mathbf{h}_i)$. Finally, the updated set-points are transmitted to the child areas in Step 5. We highlight several important aspects of this controller.

\smallskip

\begin{enumerate}
\item[(i)] \textbf{Feedback-Based Optimization:} The \ac{LC} controller directly uses real-time local measurements from the \ac{CA}. This use of feedback allows the \ac{LC} to react to unmeasured disturbances, and confers significant robustness against imperfections in knowledge of the sensitivity matrices $(\mathbf{A}_i,\mathbf{B}_i,\mathbf{M}_i,\mathbf{H}_i)$. Tuning selections which guarantee stability will be established in Section \ref{Sec:Stability}.
\item[(ii)] \textbf{Localized Control:} The design and online implementation of the controllers uses only local network information. Measurements of current, voltage, and power, local grid sensitivity matrices, and DER costs/limits are used locally within each \ac{CA} by the \ac{LC}; this information is not shared.
\item[(iii)] \textbf{Scalability:} All coordination between \acp{CA} occurs through the passing of set-points down through the feeder, from parent areas to child areas, as described in Section \ref{Sec:FeederArchitecture}. 
This minimal coordination allows the architecture to be scalable to extremely large distribution systems. While in principle this scalability should result in a decrease in speed of the overall system compared to a centralized solution, our case studies in Section \ref{sec: Simulations} indicates that this effect is indeed minor and can be overcome.
\item[(iv)] \textbf{Computational Burden:} The computation involved in Algorithm \ref{Alg:1} is dominated by the set-point update in Step 4, which requires the solution of a local convex optimization problem. This is a centralized optimization problem for the \ac{LC} to solve, which  scales with the number of \acp{DER} to be controlled, and standard methods can be applied. For instance, if $f_i$ is quadratic and the constraint set $\boldsymbol{\mathcal{X}}_i$ is polytopic, then Step 4 is a quadratic program. 
\end{enumerate}


\section{Stability Analysis and Tuning of Proposed Algorithm}
\label{Sec:Theory}

In this section we pursue a stability analysis of the proposed design. To begin, let
\begin{equation}\label{Eq:Measurements}
\mathbf{y}_i = \mathrm{col}(\mathbf{p}_{0,i},\mathbf{q}_{0,i},\mathbf{v}_i,\mathbf{i}_i)
\end{equation}
denote the vector of grid measurements taken by the $i$th \ac{LC}, which are used within Algorithm \ref{Alg:1}. Recall that the linearized distribution system model \eqref{eq:DN_Equations} captures the \emph{local} model of the grid within the $i$th \ac{CA}, and the sensitivity matrices 
\begin{equation}\label{Eq:LocalSensMatrices}
\mathbf{K}_i \define \mathrm{col}(\mathbf{M}_i,\mathbf{H}_i,\mathbf{A}_i,\mathbf{B}_i)
\end{equation}
in \eqref{eq:DN_Equations} are used within Algorithm \ref{Alg:1}. This local model ignores the impacts of \ac{DER} actions in \emph{other} \acp{CA}. In contrast, the measurements \eqref{Eq:Measurements}, as they are generated by the real grid, will include these interactions. To capture this, we introduce a full linearized model of the feeder which relates all DER set-points $\mathbf{x} = \mathrm{col}(\mathbf{x}_1,\ldots,\mathbf{x}_N)$ to all area measurements $\mathbf{y} = \mathrm{col}(\mathbf{y}_1,\ldots,\mathbf{y}_N)$, as $\mathbf{y} = \mathbf{K}\mathbf{x} + \mathbf{k}$, or
\begin{equation}\label{Eq:FullGrid}
\begin{bmatrix}
\mathbf{y}_1 \\ \vdots \\ \mathbf{y}_{N}
\end{bmatrix} = \begin{bmatrix}
\mathbf{K}_{11} & \cdots & \mathbf{K}_{1N}\\
\vdots & \ddots & \vdots\\
\mathbf{K}_{N1} & \cdots & \mathbf{K}_{NN}
\end{bmatrix}\begin{bmatrix}\mathbf{x}_{1} \\ \vdots \\ \mathbf{x}_N\end{bmatrix} + \begin{bmatrix}\mathbf{k}_1 \\ \vdots \\ \mathbf{k}_{N}\end{bmatrix}
\end{equation}
for appropriate matrices $\mathbf{K}_{ij}$ and vectors $\mathbf{k}_j$. Note that if the local sensitivity model \eqref{Eq:LocalSensMatrices} is accurate, then we expect that $\mathbf{K}_i \approx \mathbf{K}_{ii}$. The closed-loop control system now consists of Algorithm \ref{Alg:1} for each $\ac{CA}$ $i \in \{1,\ldots,N\}$ with the distribution system model \eqref{Eq:FullGrid}.


\subsection{Equilibrium Analysis}
\label{Sec:Equilibrium}

As the controller of Algorithm \ref{Alg:1} was developed beginning from the optimization problem \eqref{eq:LC-Problem}, one should expect some relationship between the equilibrium points of the closed-loop system and the optimal points of the problems \eqref{eq:LC-Problem}. We will show that the closed-loop equilibrium can be understood as the \emph{\acf{GNE}} of a related set of game-theoretic optimization problems, one for each \ac{CA}.\footnote{While the equilibrium is best understood in game-theoretic terms, this \emph{does not} imply that the \acp{CA} are in competition with one another, and indeed \acp{CA} cooperate within our scheme by accepting set-points from parents and sending set-points to children. The Nash equilibrium concept is the natural one due purely to the area-wise decentralized nature of the control system.}

When taking into account the full distribution system model \eqref{Eq:FullGrid}, the optimization problems \eqref{Eq:LCCost} should be expressed together as the set of $N$ decentralized decision problems
\begin{equation}\label{Eq:CoupledOpt}
\mathbb{P}_{i}(\mathbf{x}_{-i}): \qquad \begin{aligned}
\inf_{\mathbf{x}_i \in \boldsymbol{\mathcal{X}}_i} &\quad f_i(\mathbf{x}_i)\\
\text{s.t.} &\quad \mathbf{C}_i\mathbf{y}_i + \mathbf{b}_i + \mathbf{D}_i\mathbf{x}_{\mathsf{P}(i)} \leq 0\\
\end{aligned}
\end{equation}
for $i \in \{1,\ldots,N\}$, where $\mathbf{y}_i$ is determined by \eqref{Eq:FullGrid}, where
\begin{equation}\label{Eq:BCD}
\begin{aligned}
\mathbf{b}_i &= -\mathrm{col}(E_{i_{\rm p}}, E_{i_{\rm p}}, E_{i_{\rm q}}, E_{i_{\rm q}}, \overline{\mbf{v}}_i, -\underline{\mbf{v}}_i, \overline{\mbf{i}}_i)\\
\mathbf{C}_i &= \mathrm{blkdiag}\left(
\left[\begin{smallmatrix}s_i\vones[]^{\T}\\-s_i\vones[]^{\T}\end{smallmatrix}\right], \left[\begin{smallmatrix}s_i\vones[]^{\T}\\-s_i\vones[]^{\T}\end{smallmatrix}\right], \left[\begin{smallmatrix}I \\ -I\end{smallmatrix}\right], I\right)\\
\mathbf{D}_i &= \mathrm{col}(-s_i T_i^{\rm p},
s_i T_i^{\rm p},
-s_i T_i^{\rm q},
s_i T_i^{\rm q}, 0,0,0),
\end{aligned}
\end{equation}
In \eqref{Eq:CoupledOpt}, the decisions of the other \acp{LC} $\mathbf{x}_{-i} = (\mathbf{x}_j)_{j\neq i}$ are interpreted as fixed. A \emph{\acf{GNE}} is a collection of set-point decisions $(\mathbf{x}_1^{\star},\ldots,\mathbf{x}_N^{\star})$ such that $\mathbf{x}_i^{\star} \in \argmin_{\mathbf{x}_i} \mathbb{P}_i(\mathbf{x}_{-i}^{\star})$ for all $i \in \{1,\ldots,N\}$. Consider now the modified set of problems $\mathbb{P}_i^{\prime}(\mathbf{x}_{-i})$ given by
\begin{equation}\label{Eq:ModifiedNash}
\begin{aligned}
\inf_{\mathbf{x}_i \in \boldsymbol{\mathcal{X}}_i} \,\, f_i(\mathbf{x}_i) + \tfrac{r_i^{\rm p}}{2}\|\mathbf{x}_i\|_2^2 + \mathcal{M}_{i}(\mathbf{C}_i\mathbf{y}_i + \mathbf{b}_i + \mathbf{D}_i\mathbf{x}_{\mathsf{P}(i)})
\end{aligned}
\end{equation}
where again $\mathbf{y}_i$ is given by \eqref{Eq:FullGrid}. Compared to the problems $\mathbb{P}_i(\mathbf{x}_{-i})$, in $\mathbb{P}_i^{\prime}(\mathbf{x}_{-i})$ we have (i) introduced additional convexity into the objective function with the term $\tfrac{r_i^{\rm p}}{2}\|\mathbf{x}_i\|_2^2$, and (ii) \emph{softened} the inequality constraints by replacing them with a differentiable quadratic penalty function $\mathcal{M}_{i}$, with penalty weights given by the inverse elements of $\mathbf{R}^{\rm d}_{i}$. For instance, the first component of $\mathcal{M}_i$ is
\[
\mathcal{M}_{i,1}(\xi_{i,1}) = \begin{cases}
0, &\xi_{i,1} \leq 0\\
\frac{1}{2r_{\lambda_i}}|\xi_{i,1}|^2, &\xi_{i,1} > 0,
\end{cases}
\]
When both $r_{i}^{\rm p}$ and $\mathbf{R}^{\rm d}_{i}$ are small, the  problem \eqref{Eq:ModifiedNash} closely approximates the problem \eqref{Eq:CoupledOpt}. While the details are beyond our scope here, if $\mathbf{K}_i = \mathbf{K}_{ii}$, then Algorithm \ref{Alg:1} is precisely a measurement-based and decentralized dual gradient algorithm for computing a \ac{GNE} of \eqref{Eq:ModifiedNash}; see \cite{0158_Agarwal2022, 0159_PavelBook2012, GB-DLM-MHDB-SB-JL-FD:21} for related game-theoretic online optimization concepts. In summary, and in rough terms, this means that each \ac{LC} will make the best set-point decision that it can, given the limited information it has due to the decentralized control architecture.


\subsection{Closed-Loop Stability Analysis}
\label{Sec:Stability}

Section \ref{Sec:Equilibrium} interprets closed-loop equilibrium points as generalized Nash equilibria, but does not assert that an equilibrium point exists, nor whether it is stable; our main stability result addresses both of these items.

As notation, let $\boldsymbol{\alpha} = \mathrm{blkdiag}(\boldsymbol{\alpha}_1,\ldots,\boldsymbol{\alpha}_N)$ and similarly for block diagonal $\mathbf{C}, \mathbf{D}, \mathbf{R}^{\rm d}$, and let $\mathbf{K}_{\rm d} = \mathrm{blkdiag}(\mathbf{K}_1,\ldots,\mathbf{K}_N)$. Recall the adjacency matrix $\mathcal{A}$ of Section \ref{Sec:FeederArchitecture}, and let $\boldsymbol{\mathcal{A}}$ denote an expanded version of this matrix, where each $1$ or $0$ becomes an identity matrix or zero matrix of appropriate dimension. Based on these, and on the previously defined control parameters, define the constant $\mathsf{L} > 0$ and the matrix $\mathsf{M} \in \real^{N \times N}$ by
\[
\begin{aligned}
 \mathsf{L} &= \frac{\|\mathbf{R}^{\rm d}\|_2 + \|\mathbf{C}\mathbf{K} + \mathbf{D}\boldsymbol{\mathcal{A}}^{\T}\|_2 \|\mathbf{K}_{\rm d}\|_2\|\mathbf{C}\|_2}{\min_{i\in\{1,\ldots,N\}}(m_i + r_{i}^{\rm p})}\\
\mathsf{M}_{ij} &= \begin{cases}
\|\mathbf{R}_{i}^{\rm d}\|_2 - \frac{\|\mathbf{K}_{ii} - \mathbf{K}_i\|_{2}}{m_i + r^{\rm p}_{i}} (\|\mathbf{D}_i\|_2\|\mathbf{C}_i\|_2^2\|\mathbf{K}_{i}\|_2), &i = j\\
-\frac{\|\mathbf{K}_{ij}\|_2}{m_j+r^{\rm p}_{j}}(\|\mathbf{C}_i\|_2\|\mathbf{C}_j\|_2 \|\mathbf{K}_{j}\|_2) - N_{ij},& i \neq j
\end{cases}
\end{aligned}
\]
where $N_{ij} = \mathcal{A}_{ij}\|\mathbf{D}_i\|_2\|\mathbf{C}_j\|_2/(m_j + r^{\rm p}_{j})$. We can now succinctly state the main result.

\medskip

\begin{theorem}[\bf Closed-Loop Stability]\label{Thm:Stability}
Consider the closed-loop system consisting of Algorithm \ref{Alg:1} for each $\ac{CA}$ $i \in \{1,\ldots,N\}$ with the distribution system model \eqref{Eq:FullGrid}.  If $\mathsf{M} + \mathsf{M}^{\T} \succ 0$, then the closed-loop system possess a unique equilibrium point $(\mathbf{x}_i^{\star},\mathbf{d}_i^{\star})_{i\in\{1,\ldots,N\}}$, and the equilibrium is globally exponentially stable for all gain selections $\boldsymbol{\alpha} = \mathrm{blkdiag}(\boldsymbol{\alpha}_1,\ldots,\boldsymbol{\alpha}_N)$ satisfying $\lambda_{\rm max}(\boldsymbol{\alpha})^2/\lambda_{\rm min}(\boldsymbol{\alpha}) < \lambda_{\rm min}(\mathsf{M} + \mathsf{M}^{\T})/\mathsf{L}^2$.
\end{theorem}

\medskip

The proof can be found in the Appendix. The diagonal elements of the matrix $\mathsf{M}$ can be interpreted as capturing the margin of ``local'' closed-loop stability for each $\ac{CA}$, while the off-diagonal elements capture any potentially negative effects of interaction between the \acp{CA}. The stability condition $\mathsf{M} + \mathsf{M}^{\T} \succ 0$ then has the elegant interpretation that local stability should outweigh inter-area coupling. It is clear that this condition can \emph{always} be satisfied by selecting sufficiently large values for the dual and primal regularization parameters $\mathbf{R}^{\rm d}_{i}$ and $r_{i}^{\rm p}$, and that  smaller values for these parameters are permissible if the local sensitivity mismatch $\|\mathbf{K}_{ii} - \mathbf{K}_i\|_{2}$ and the cross-area coupling $\|\mathbf{K}_{ij}\|_2$ are small. The gain restriction $\lambda_{\rm max}(\boldsymbol{\alpha})^2/\lambda_{\rm min}(\boldsymbol{\alpha}) < \lambda_{\rm min}(\mathsf{M} + \mathsf{M}^{\T})/L^2$ states that one can obtain a stable tuning by starting $\boldsymbol{\alpha}$ small and slowly increasing. Further details on tuning will be presented next.


\subsection{Practical Tuning Guidelines}
\label{Sec:Tuning}

Before moving to our case studies, we provide practical guidelines for tuning the parameters in Algorithm \ref{Alg:1}. Each \ac{LC} must set the following parameters, which can be systematically tuned and set, as follows:



\smallskip

\subsubsection{Sampling period $T_{\rm s}$} The sampling period is mainly constrained by the quality and speed of the communication infrastructure; see \cite{0160_DILEEP2020,0161_ALI2017} for discussion on communication technologies and standards. As is the case in all digital control systems, lower sampling periods are preferred. 


\smallskip

\subsubsection{Cost functions $f_{ij}(x_j)$}\label{sec: tuning - cost function section}

The \acp{DER} and \acp{VDER} cost functions will determine the relative steady-state allocation of control actions to \acp{DER} by the \ac{LC}. The \ac{LC} manager can set these costs to preferentially use or discourage certain \acp{DER} based on any desired operational criteria, e.g., speed of response, or to introduce different marginal costs for high vs. low utilization of a \ac{DER}. As mentioned previously, for \acp{VDER} the selection \eqref{eq:Cost of VDER} will mimic the case of a single grid-wide centralized dispatch.

%


\subsubsection{Tracking Tolerances $E_{i_{\rm p}}$ and $ E_{i_{\rm q}}$} These tolerances should be set based on desired set-point tracking accuracy at the feeder head. Due to the constraint-softening effects of regularization (Section \ref{Sec:Equilibrium}), tightening of these tolerances may be beneficial.

\subsubsection{Regularization parameters $r_{i}^{\rm p}$ and $\mbf{R}_{i}^{\rm d}$}
From the result of Theorem \ref{Thm:Stability}, larger values of regularization parameters help ensure closed-loop stability. Conversely though, from the discussion in Section \ref{Sec:Equilibrium}, larger regularization parameters lead to softer enforcement of voltage, current, and tracking constraints. Thus, there is a trade-off; these parameters should be large enough to ensure stability, but small enough to ensure minimal or no constraint violation. From \eqref{Eq:Rdi}, setting $\mbf{R}_{i}^{\rm d}$ means setting 7 parameters for each \ac{CA}. To simplify this, we express these 7 parameters as multiples of a single constant $\tilde{r}_i^{\rm d} > 0$, as shown in the third column of Table \ref{table: Duals steps and rds}; the constants $(c_{\lambda,i}, c_{\mu,i},\ldots)$ are unit conversions, and are shown in the third column of Table \ref{table: LCs Config}. For each \ac{LC}, the only regularization parameters to set are now $\tilde{r}_i^{\rm d}$ and $r_i^{\rm p}$; these values can always be initialized for stability based on Theorem \ref{Thm:Stability}, and then decreased if voltage/current constraint violation is observed.

\subsubsection{Dual Step Sizes $\bs{\alpha}$} The step sizes $\bs{\alpha}$ control how aggressively each \ac{LC}  reacts to constraint violations, and can be thought of as integral control gains. From \eqref{Eq:stepSizes}, each \ac{LC} has 7 such gains to set, and it is again helpful to express all gains as multiples of a single dimensionless constant $\alpha_i > 0$, as shown in the second column of Table \ref{table: Duals steps and rds}; the constants $(a_{\lambda,i}, a_{\mu,i}, \ldots)$ are shown in the second column of Table \ref{table: LCs Config}. This reduces the gain tuning to a single parameter $\alpha_i$; following Theorem \ref{Thm:Stability}, we recommend that one slowly increases $\alpha_i$ to ensure stability. The constants $(a_{\lambda,i}, a_{\mu,i}, \ldots)$ are fixed based on unit conversions, but are further adjusted to reflect the relative dynamic importance of voltage/current limits, and power tracking constraints in implementation (see Sections \ref{Sec:5BusFeeder}-\ref{sec: IEEE-8500}). In particular

\begin{enumerate}[(i)]
    \item the constants ($a_{\bs{\gamma},i}$, $a_{\bs{\nu},i}$) associated with voltage constraints have been made larger compared to constant ($a_{\bs{\zeta},i}$) associated with current constraints; this ensures voltage constraints are quickly maintained, as transient violation of currents above their steady-state limits is acceptable over time-frames of $\sim 10-20$s.

    \item the constants ($a_{\lambda,i}$, $a_{\mu,i}$) associated with active power tracking have been made larger compared to the constants ($a_{\eta,i}$, $a_{\psi,i}$) associated with reactive power tracking. This prioritizes fast active power tracking, and minimizes transient voltage fluctuations, particularly in \acp{CA} deeper within the network.
\end{enumerate}

\begin{table}[ht!]
  \centering
  \caption{Dual variables step-sizes and regularization parameters; units of the scaling coefficients can be inferred from Algorithm \ref{Alg:1}.}
  \label{table: Duals steps and rds}
  \renewcommand{\arraystretch}{1.1} 
  \begin{tabular}{ccc}
    \toprule
    Controller State (unit) & Gain & Regularization \\
    \midrule
    $\lambda_i \ (\mathrm{W})$ & $\alpha_{\lambda,i} = a_{\lambda,i}\alpha_i$ & $r_{\lambda,i} =  c_{\lambda,i}\tilde{r}^{\rm d}_i$ \\
    $\mu_i \ (\mathrm{W})$ & $\alpha_{\mu,i} = a_{\mu,i}\alpha_i$ & $r_{\mu,i} =  c_{\mu,i}\tilde{r}^{\rm d}_i$ \\
    $\eta_i \ (\mathrm{Var})$ & $\alpha_{\eta,i} = a_{\eta,i}\alpha_i$ & $r_{\eta,i} =  c_{\eta,i}\tilde{r}^{\rm d}_i$ \\
    $\psi_i \ (\mathrm{Var})$ & $\alpha_{\psi,i} = a_{\psi,i}\alpha_i$ & $r_{\psi,i} =  c_{\psi,i}\tilde{r}^{\rm d}_i$ \\
    $\bs{\gamma}_i \ (\frac{\mathrm{W}^2}{\mathrm{V}})$ & $\alpha_{\bs{\gamma},i} = a_{\bs{\gamma},i}\alpha_i$ & $r_{\bs{\gamma},i} =  c_{\bs{\gamma},i}\tilde{r}^{\rm d}_i$ \\
    $\bs{\nu}_i \ (\frac{\mathrm{W}^2}{\mathrm{V}})$ & $\alpha_{\bs{\nu},i} = a_{\bs{\nu},i}\alpha_i$ & $r_{\bs{\nu},i} =  c_{\bs{\nu},i}\tilde{r}^{\rm d}_i$ \\
    $\bs{\zeta}_i \ (\frac{\mathrm{W}^2}{\mathrm{A}})$ & $\alpha_{\bs{\zeta},i} = a_{\bs{\zeta},i}\alpha_i$ & $r_{\bs{\zeta},i} =  c_{\bs{\zeta},i}\tilde{r}^{\rm d}_i$ \\
    \bottomrule
  \end{tabular}
\end{table}


\begin{table}[ht!]
  \centering
  \caption{Default configurations for \ac{LC} controllers.}
  \label{table: LCs Config}
  \renewcommand{\arraystretch}{1.1} 
  \setlength{\tabcolsep}{3pt}
  \begin{tabular}{cc|cc|cc}
    \toprule
    Parameter & Value & Parameter & Value & Parameter & Value \\
    \midrule
    $\alpha_i$          &   $0.002$             &   $a_{\lambda,i}$     &   $10^{3}$    &   $c_{\lambda,i}$     &   $10^{-3}$   \\
    $r^{\rm p}_i$       &   $10^{-4}$       &   $a_{\mu,i}$         &   $10^{3}$    &   $c_{\mu,i}$         &   $10^{-3}$   \\
    $\tilde{r}^{\rm d}_i$     &   $10^{-3}$       &   $a_{\eta,i}$        &   $10^{3}$    &   $c_{\eta,i}$        &   $10^{-3}$   \\
    $E_{i_{\rm p}}$           &   $100$W       &   $a_{\psi,i}$        &   $10^{3}$    &   $c_{\psi,i}$        &   $10^{-3}$   \\
    $E_{i_{\rm q}}$           &   $100$Var       &   $a_{\bs{\gamma},i}$ &   $10^{12} \frac{\mathrm{W}^2}{\mathrm{V}^2}$   &   $c_{\bs{\gamma},i}$ &   $10^{-12}\frac{\mathrm{V}^2}{\mathrm{W}^2}$  \\
    $\overline{\mbf{v}}_i$    &   $1.05$p.u.   &   $a_{\bs{\nu},i}$    &   $10^{12} \frac{\mathrm{W}^2}{\mathrm{V}^2}$   &   $c_{\bs{\nu},i}$    &   $10^{-12}\frac{\mathrm{V}^2}{\mathrm{W}^2}$  \\
    $\underline{\mbf{v}}_i$   &   $0.95$p.u.   &   $a_{\bs{\zeta},i}$  &   $10^{7} \frac{\mathrm{W}^2}{\mathrm{A}^2}$    &   $c_{\bs{\zeta},i}$  &   $10^{-7}\frac{\mathrm{A}^2}{\mathrm{W}^2}$   \\
    \bottomrule
  \end{tabular}
\end{table}

\section{Case Studies}\label{sec: Simulations}

We present three case studies of increasing complexity to illustrate and validate the proposed design: (1) a simple 5 bus feeder, (2) the IEEE-123 bus feeder, and (3) the IEEE-8500 bus feeder. The simple 5-bus feeder will be used to demonstrate the basic functionality of the controller, including how \acp{CA} and \acp{LC} interact with one another; the latter two test systems will demonstrate scalability of the approach.

The tests were run using a customized application, MATDSS, that was developed and run using MATLAB\textsuperscript{\textregistered} 2023a and OpenDSS\textsuperscript{\copyright} 9.6.1.3; this software has been made available at \cite{Farhat_MATDSSApplication}. The feeders are defined using OpenDSS Scripting Language, with the IEEE-123 and IEEE-8500 feeders receiving minor modifications from their original versions in OpenDSS\textsuperscript{\copyright}; see  \cite{FarhatSupplOnlineMat} for details on these modifications. We set all \acp{LC} control parameters to the nominal values in Table \ref{table: LCs Config}, unless specified otherwise; current limits vary by line, and are omitted due to space limitations. Controllable \acp{DER} are integrated throughout the test systems. As our only requirement is that these \acp{DER} are responsive to dispatch commands, the internal dynamics and specific nature of the \acp{DER} are of secondary importance; each \ac{DER} is modelled as having a first-order response to power commands with time constant $\tau$; more detailed \ac{DER} models with internal controls can be easily integrated within MATDSS, see \cite{FarhatSupplOnlineMat, Farhat_MATDSSApplication}. All \ac{DER} power limits are box constraints $\mathcal{X}_j = [\underline{x}_j,\overline{x}_j]$; see Table \ref{table: DERs Config}.

\begin{table}[!htb]
  \centering
  \caption{DER's Default parameters.}
  \label{table: DERs Config}
  \renewcommand{\arraystretch}{1.4} 
  \begin{tabular}{cccc}
    \toprule
    Parameter & Value & Parameter & Value\\
    \midrule
    $\tau_j$    &   $0.2$s     & $C^{\prime\prime}_j$, $C^{\prime}_j$&   $\mathrm{diag}(20,20)$, $(0,0)$ \\
    $\underline{x}_j$ &   $(-10^{6} \mathrm{W}, -10^{6}\mathrm{Var})$    &   $\overline{x}_j$    &   $(10^{6} \mathrm{W}, 10^{6}\mathrm{Var})$\\
    \bottomrule
  \end{tabular}
\end{table}

Each test system is divided into \acp{CA}. Our design will be compared and contrasted with a baseline centralized controller, labelled $1$\ac{CA}, which is a 1-area implementation of our hierarchical controller, acting with global information to control \emph{all} \acp{DER} within the feeder; this serves as a ``best case'' against which to compare our hierarchical design. For all feeders, we consider the case were the \ac{TSO} request is of the form $X_{0}^{\rm set} = (\mathrm{p}_0^{\rm set}, 0)$, i.e., active power tracking at the \ac{DN}-\ac{TN} interface. The sampling time of the controller is set as $T_{\rm s} = 100$ms. Throughout the tests and for all controllers, we set $r^{\rm p}_i = 10^{-4}$, $\tilde{r}_i^{\rm d} = 10^{-3}$, and $\alpha_i = 0.002$, and set the corresponding gains as in Table \ref{table: LCs Config}. With these regularization parameters, for tracking within $\pm 1$kW for $\mathrm{p}_{i,{0}}^{\rm set}$, we tighten the power tracking constraints and set $E_{i_{\rm p}} = 100$W and $E_{i_{\rm q}} = 100$Var.







\subsection{5-Bus Feeder}
\label{Sec:5BusFeeder}

\begin{figure}[ht!]
    \centering
    \includegraphics[width=0.35\textwidth]{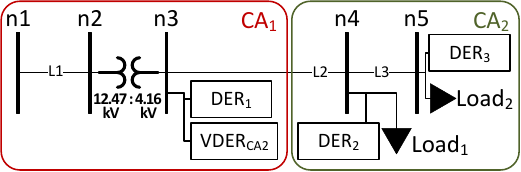}
    \caption{Three phase 5-Bus feeder circuit.}\label{fig:5 bus feeder circuit}
\end{figure}

Consider the three-phase 5-bus feeder of Figure \ref{fig:5 bus feeder circuit}, which has been partitioned into two \acp{CA}. Within the feeder, three \acp{DER} are placed at buses n$3$, n$4$ and n$5$, with a \ac{VDER} added to \ac{CA}1, representing the \ac{LC} in \ac{CA}2. There are two loads of $400$kW and $770$kW with power factor of $0.9$ at buses n$4$ and n$5$, respectively. The purpose of our test here is to illustrate the basic behaviour and response of the controller.

 

\begin{figure}[ht!]
    \centering
   
    \includegraphics[width=0.45\textwidth]{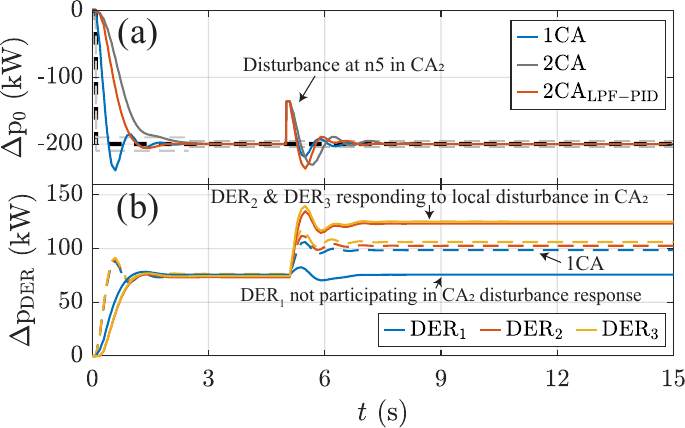}
    {\phantomsubcaption\label{fig:5BusStepTracking_P0}%
    \phantomsubcaption\label{fig:5BusStepTracking_DERCurves}}%
    \caption{5-Bus feeder step-tracking with disturbance response. (a) Tracking of $\Delta \mathrm{p}_0$ with 1\ac{CA} (blue) and 2\ac{CA} (orange) configurations. (b) \acp{DER} active power responses. Dashed lines corresponds to single-area (1CA) while solid lines corresponds to multi-area ($2 \rm{CAs}_{\rm LPF-PID}$). {(1\ac{CA}: $a_{\lambda,1},a_{\mu,1} = 5\times 10^3$, 2\acp{CA}: $a_{\lambda,2}, a_{\mu,2} = 5\times 10^3$, 2${\rm \acp{CA}}_{\rm LPF-PID}$: $\alpha_1, \alpha_2 = 0.003$, $a_{\lambda,2}, a_{\mu,2} = 5\times 10^3$)}}\label{fig:5BusStepTracking}
\end{figure}

A reference change of $200$kW is requested at the feeder head at $t = 0$, and a load change (disturbance) of $100$kW ($0.9$pf) at $t = 5$s located at bus n$5$. The change in power at the feeder head serves as a metric to gauge the tracking performance of the controllers. Simultaneously, the response to disturbances provides insights into how both centralized and decentralized controllers effectively manage local constraints and disturbances. The power response at the feeder head is shown in Figure \ref{fig:5BusStepTracking_P0}, which as discussed plots both a hierarchical two-\acp{CA} and a centralized one-\ac{CA} implementation. The figure shows the change in power flow at the \emph{interface bus} of the feeder head across all phases, i.e. $\Delta {\rm p}_0 = \Delta \vones^\T \mathbf{p}_{1,0}$.

The multi-area implementation (2\acp{CA}) shows a more sluggish response compared to the centralized implementation (1\ac{CA}). While not covered by our theory, we have found this sluggishness can be overcome by (i) incorporating proportional and derivative (PD) control action into the \ac{LC} controller, and (ii) passing the control signal for \acp{VDER} through a \ac{LPF}. To explain the first modification, consider the $\lambda_i$ update in Algorithm \ref{Alg:1}. This can be viewed as integral-type controller $\lambda_i^+ = \mathcal{P}_{\geq 0}(\lambda_i + \alpha_{\lambda_i}e_{\lambda_i})$ acting on the error $e_{\lambda_i} \coloneqq \vones^\T \mathbf{p}_{i,0} - \mathrm{p}_{i,{0}}^{\rm set} - E_{i_{\rm p}} - r_{\lambda_i} \lambda_i$. Let $\mathbf{e}_i$ denote the stacked vector of all these errors. Incorporating PD action alongside this integral controller can accelerate the overall response time. This is done by modifying Step 4 in Algorithm \ref{Alg:1}, wherein the argument $\mathbf{d}_i^+$ is replaced by $\tilde{\mathbf{d}}_i^+$, where
\[
\tilde{\mathbf{d}}_i^+ = \mathbf{d}_i^+ + \kappa_{\rm p}\mathbf{e}_i + \kappa_{\rm d}(\mathbf{y}_i - \mathbf{y}_i^-),
\]
%
%
where $\mathbf{y}_i = \mathrm{col}(\vones^\T \mathbf{p}_{i,0}, -\vones^\T \mathbf{p}_{i,0}, \vones^\T \mathbf{q}_{i,0}, -\vones^\T \mathbf{q}_{i,0}, \mathbf{v}_i, -\mathbf{v}_i, \mathbf{i}_i)$ is the stacked vector of raw measurements, and $\kappa_{\mathrm{p}}$ and $\kappa_{\rm d}$ are diagonal matrices of proportional and derivative gains. We have found that derivative action need only be used for VDERs. The second modification, a low-pass filter, is used to eliminate the passing of aggressive control actions down through VDERs, which allows higher level controllers to use larger integral gains. After Step 4 
 of Algorithm \ref{Alg:1}, the VDER set-points are passed through a low-pass filter before being sent to the child areas. See the supplement \cite{FarhatSupplOnlineMat} for details on tuning of the proportional-derivative gains and low-pass filter time constants.

Returning now to Figure \ref{fig:5BusStepTracking}, the $\rm 2CA_{LPF-PID}$ curve shows response of the 2CAs structure with PID controllers and LPF filtering. With this implementation, one can achieve tracking results similar to single-area system (1CA) in terms of power set-points while preserving data privacy and maintaining a hierarchical control structure. The settling time after the step change was $1.07$s for 1CA, $1.78$s for 2CA, and $1.02$s for 2CA with PID controllers and LPF filter. Note that the disturbance at $t =5$s, which occurs in the child area, is quickly rejected by both the 1CA and 2CAs implementations. 
 
 Figure \ref{fig:5BusStepTracking_DERCurves} plots the active power responses of the three \acp{DER} during the test. 
Notably, both 1CA and 2CAs ($\rm LPF-PID$) implementations exhibited similar settling times and \ac{DER} participation when responding to the initial step-change. Importantly however, when compensating for the disturbance at $t=5$s located at bus n$5$, the 1CA and 2CA implementations behave differently. In the 2CAs implementation, the disturbance was regarded as local perturbation within \ac{CA}2, and consequently, only $\text{\ac{DER}}_2$ and $\text{\ac{DER}}_3$ were responsible for providing compensation. In contrast, the 1CA implementation redispatched all \acp{DER} to mitigate the disturbance. In subsequent plots and tests, all multi-CA implementations will include PID action and low-pass filters, and we drop the `$\rm LPF-PID$' annotation.


\begin{figure}[ht!]
    \centering
    \includegraphics[width=0.45\textwidth]{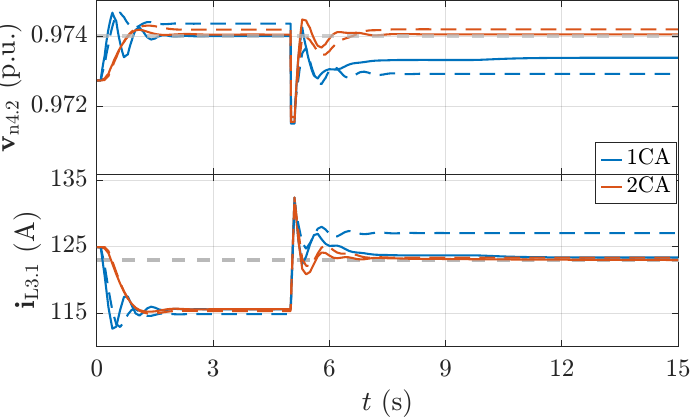}
    \caption{5-Bus feeder step-tracking with disturbance voltage and current control with 1\ac{CA} (blue) and 2\ac{CA} (orange) configurations. Dashed lines corresponds to regular constraints on voltage ($\overline{\mbf{v}} = 1.05$ p.u.) and $135$A current limit. }\label{fig:5BusStepTracking_IVControl}
\end{figure}

The dashed lines in Figure \ref{fig:5BusStepTracking_IVControl} show the voltage at bus n4 and the current on line L3 during the test; with voltage limits set to their nominal values and current limit of $135$A. To illustrate how effectively the controllers maintain local circuit constraints, we tighten the voltage and current constraints to $\overline{\mathbf{v}} = 0.974$p.u. and $\overline{\mathbf{i}} = 123$A, and repeat the test, with results plotted in solid lines. Both configurations, 1\ac{CA} and 2\ac{CA}, showed similar response and enforced the operational constraints in equilibrium. For details on the controller gain settings, please refer to the supplement \cite{FarhatSupplOnlineMat}. 


\subsection{IEEE-123 Feeder}\label{sec: IEEE-123}
\begin{figure}[!ht]
    \centering
    \includegraphics[width=0.40\textwidth]{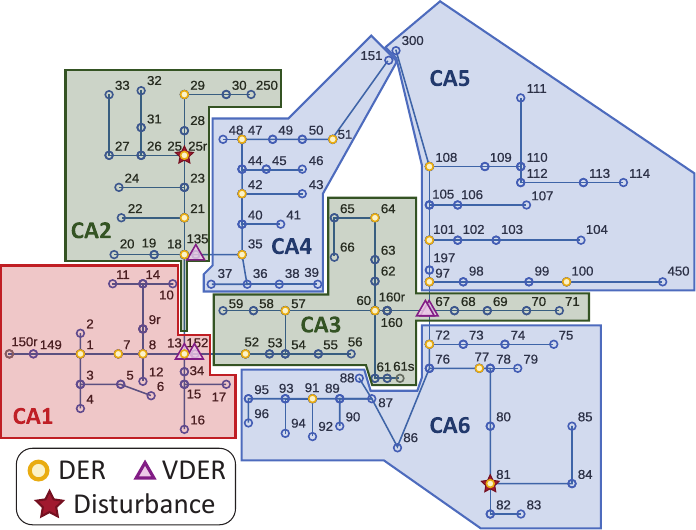}
    \caption{IEEE-123 bus feeder with six control areas.}\label{fig:IEEE-123 feeder circuit}
\end{figure}

Following the proposed control structure described in Section \ref{sec: Multi-level Control Structure}, the IEEE-123 feeder is partitioned into 6 control areas as shown in Figure \ref{fig:IEEE-123 feeder circuit}. The feeder head area (CA1) is the parent area for both CA2 and CA3, with CA4 a further child of CA2, and CA5 and CA6 further children of CA3. This permits an investigation of the impact of multiple parent-child layers on the control performance.

As shown in Figure \ref{fig:IEEE-123 feeder circuit}, there are $4$ \acp{DER} per \ac{CA}, with cost parameters  $C_{j}^{\prime\prime} = \mathrm{diag}(40, 40)$ for all \acp{DER}, and remaining parameters as in Table \ref{table: DERs Config}. All VDER costs are set according to \eqref{eq:Cost of VDER}. 
We next demonstrate the response of both configurations (1CA and 6CAs) to two test scenarios: 1) step-tracking with disturbances, and 2) stepped-ramp-tracking with disturbances.

\subsubsection{Step-tracking with disturbances}
\begin{figure}[!h]
    \centering
    \includegraphics[width=0.45\textwidth]{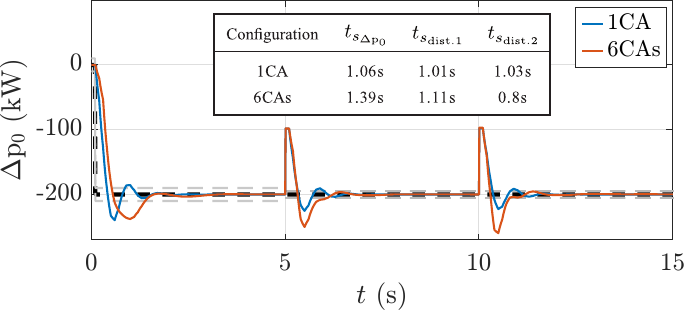}
    \caption{IEEE-123 feeder step-tracking with two disturbances. {(1CA: $\alpha_1 = 0.001$, $a_{\lambda,1}, a_{\mu,1} = 2\times 10^3$, 6CAs: $\alpha_1 = 0.0096$, $\alpha_{2,3,4,5,6} = 0.00432$, $a_{\lambda,1}, a_{\mu,1} = 5\times 10^3$)}}\label{fig:IEEE-123 Step Tracking}
\end{figure}

A step change of $200$kW is requested at the feeder head at $t = 0$s, followed by two disturbances: a $100$kW ($0.9$pf) load change at bus $25$ in \ac{CA}2 at $t = 5$s, and a $100$kW ($0.9$pf) load change at bus $81$ in \ac{CA}6 at $t = 10$s. The response of the multi-\acp{CA} centralized implementations is shown in Figure \ref{fig:IEEE-123 Step Tracking}. Note that the 6CA implementation incurs only a minor hit in performance, despite its decentralized hierarchical nature.
%
%

\subsubsection{Stepped-ramp-tracking with disturbances}
\begin{figure}[!h]
    \centering
    \includegraphics[width=0.45\textwidth]{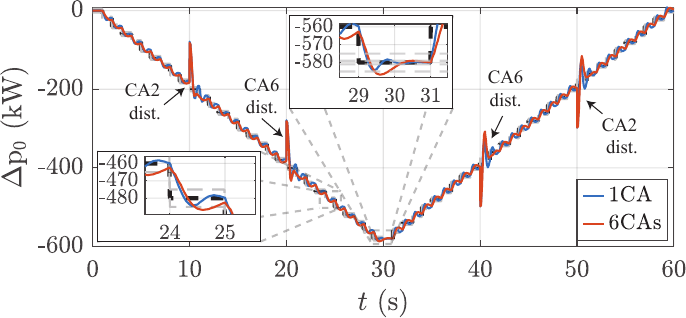}
    \caption{IEEE-123 feeder ramp-tracking with two disturbances.}\label{fig:IEEE-123 ramp Tracking}
\end{figure}

To demonstrate the tracking capabilities of the multi-CA design, we consider ramp-like reference signal at the feeder head, where a change of $20$kW occurs every $1$s. We consider the same two disturbances as in step-tracking test, where the disturbances now are triggered at different times: \emph{dist.} 1  (disturbance at bus $25$ in CA2) is connected from $t = 10$s until $t = 50$s, while \emph{dist.} 2 (disturbance at bus $81$ in CA6) is connected from $t = 20$s until $t = 40$s. The power tracking response is shown in Figure \ref{fig:IEEE-123 ramp Tracking}, with insets showing details of the transient response. The multi-\acp{CA} implementation produces similar results to the centralized 1CA implementation. 

\begin{figure}[!h]
    \centering
    \includegraphics[width=0.45\textwidth]{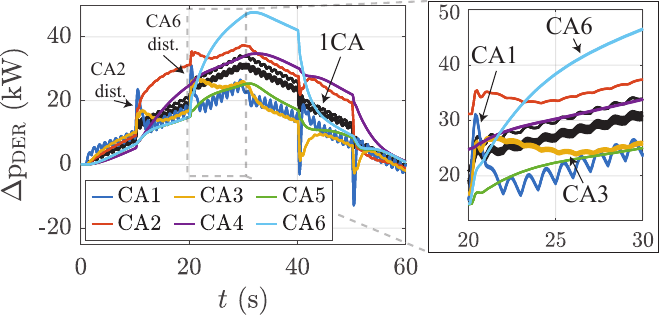}
    \caption{IEEE-123 feeder ramp-tracking \acp{DER} response. Black curves are  \acp{DER} responses for centralized 1CA implementation.}\label{fig:IEEE-123 ramp Tracking DERs response}
\end{figure}

Figure \ref{fig:IEEE-123 ramp Tracking DERs response} shows the active power responses of the \acp{DER} in the feeder, where \acp{DER} have been grouped based on the \ac{CA} they belong to, to focus on the collective response of different \acp{CA}. For the centralized controller (black curves), all \acp{DER} throughout the system show similar behavior when responding to set-point changes or disturbances, as they all have the same cost function, where all \acp{DER} respond promptly.

The \acp{DER} in the 6\ac{CA} implementation behave differently. Focusing on the first 10 seconds, after each step-change in set-point at the feeder head, \acp{DER} within \ac{CA}1 (dark blue) respond the fastest, followed by lower-level areas gradually increasing their participation. The \acp{DER} in child areas of \ac{CA}1 (i.e., \ac{CA}2 and \ac{CA}3) accelerate their response faster than those in grandchild areas. Importantly, the ``oscillations'' here are \emph{not} a form of instability, but are the result of CA1 responding aggressively to meet the set-point, then ramping down as other CAs begin to contribute. Focusing now on the disturbance at bus $81$ within \ac{CA}6 at $t = 20$s, similar observations hold (Figure \ref{fig:IEEE-123 ramp Tracking DERs response}, inset). The parent areas \ac{CA}3 and \ac{CA}1 (parent of \ac{CA}3) initially respond to maintain tracking at the feeder head, while concurrently, the contingent area (\ac{CA}6) ramps up its \acp{DER} to counteract the disturbance locally; \ac{CA}2 and \ac{CA}4 display minimal response. Thus, in the multi-area setup, local disturbances are compensated by local \acp{DER}, while parent areas ensure set-point tracking during the transient adjustment. 


%
%

\begin{figure}[!h]
    \centering
    \includegraphics[width=1\columnwidth]{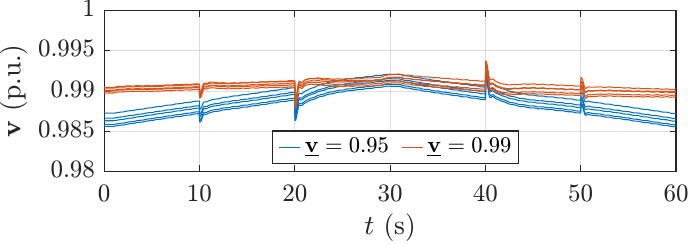}
    \caption{Voltage magnitudes for a representative set of buses using multi-\acp{CA} configuration. Blue: $\underline{\mathbf{v}} = 0.95$p.u. and Orange: $\underline{\mathbf{v}} = 0.99$p.u.}\label{fig:IEEE-123 ramp Tracking Voltage Control}
\end{figure}

Figure \ref{fig:IEEE-123 ramp Tracking Voltage Control} plots selected voltages in the circuit during the previous test, with original voltage limits (blue) and a tightened lower voltage limit $\underline{\mbf{v}} = 0.99$p.u. (orange). The controllers effectively maintain voltage levels within these new limits while responding to step-changes and disturbances. 

\subsection{IEEE-8500 Feeder}\label{sec: IEEE-8500}

Our case study of the IEEE-123 bus feeder in Section \ref{sec: IEEE-123} highlighted interactions between parent \acp{CA} and their children in response to set-point changes and disturbances. In a system of this size, centralized control may still be feasible. For very large systems however, a centralized control approach becomes cumbersome, and suffers from increased computational, communication, and information privacy issues. Our final case study on the large IEEE 8500 bus feeder is aimed at demonstrating scalability of the proposed multi-area controller, wherein the use of primarily local measurements and model information helps in overcoming the practical limitations of centralized optimization-based control.

\begin{figure}[!h]
    \centering
    \includegraphics[width=0.47\textwidth]{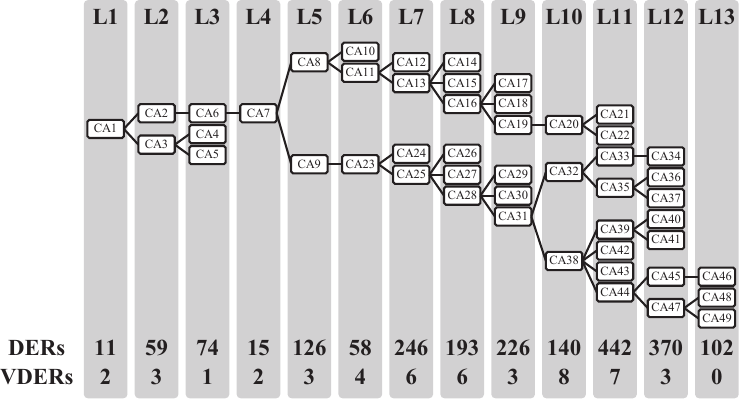}
    \caption{IEEE-8500 feeder control areas' tree with 13 nested levels of control areas.}\label{fig:IEEE-8500 Tree Node}
\end{figure}

The 8500 bus feeder is partitioned here into 49 \acp{CA} of varying sizes and composition, spread across a 13 layer hierarchy; the control area graph $\mathcal{G}_{\rm CA}$ is shown in Figure \ref{fig:IEEE-8500 Tree Node}). The areas range from 10 to 322 buses, with some areas deeper within the feeder hierarchy containing entirely single-phase circuits. This variation in area sizes demonstrates the capability of the proposed structure to effectively coordinate multiple \acp{CA} with disparate sizes and structures. \acp{DER} are placed throughout the network across all areas, with the number of \acp{DER} ranging from 4 to 127 per area, resulting in a total of 2,062 \acp{DER}, including 1-, 2-, and 3-phase \acp{DER}. For more detailed information regarding the control areas and controller configurations, please refer to the supplement \cite{FarhatSupplOnlineMat}.

\begin{figure}[!t]
    \centering
    \includegraphics[width=1\columnwidth]{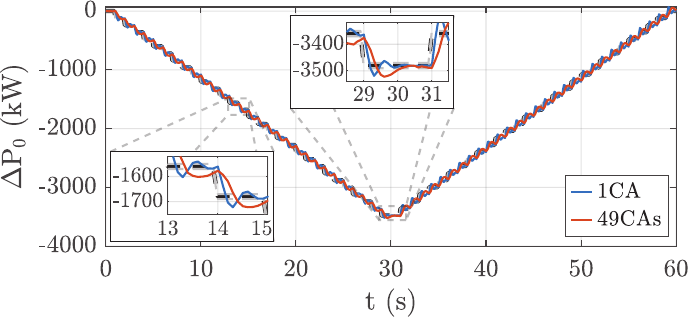}
    \caption{IEEE-8500 feeder ramp-tracking. {(1\ac{CA}: $\alpha_1 = 0.0015$, $a_{\lambda, \mu, \eta, \psi} = 20$, 49\acp{CA}: $\alpha_1 = 0.0004, \alpha_{\rm rest} = 0.0002$, $a_{\lambda,1}, a_{\mu, 1} = 5\times 10^3$)}}\label{fig:IEEE-8500 ramp Tracking}
\end{figure}

Figure \ref{fig:IEEE-8500 ramp Tracking}, plots the tracking response of both configurations, 1\ac{CA} and 49\acp{CA}, to a ramp change in power set-point at the feeder head, in increments of $120$kW each second. Despite its highly decentralized and hierarchical architecture, our multi-area control scheme produces a response similar to an ideal centralized implementation. The multi-area architecture enables privacy preservation and operational boundaries, while optimizing of thousands of \acp{DER} responses in real-time for fast \ac{TN}-\ac{DN} coordination.


\section{Conclusions}\label{sec: Conclusions}

We have developed, theoretically analyzed, and tested a multi-area feedback-based hierarchical control framework to coordinate distribution level \acp{DER} and \ac{DER} aggregators in response to \ac{TSO} power requests. The proposed framework addresses the pressing need for an effective and practical \ac{DER} coordination framework that meets four critical requirements: speed, model independence, primarily local measurement and data reliance, and privacy/operational boundary preservation. The case studies illustrate how the proposed architecture addresses the information, communication, computational, and scalability issues of a centralized feedback-optimization controller, with no significant degradation in dynamic performance compared to an idealized centralized implementation. 



There are several avenues for continued work. One drawback of the multi-area architecture is that each area controller has many parameters to tune, hence our procedure in Section \ref{Sec:Tuning} for reducing the number of tuning parameters. An important direction for future research is automatic-tuning of controllers' parameters and \acp{DER}' cost functions to achieve a desired time-domain response \cite{KJA-TH:06, VH-LH-XH-EPA-FD:23,JC-Jw-AB:23}. Another open direction is to explicitly account for different DER dynamics in the design, which would help to coordinate both fast and slow resources. Finally, we are proceeding with integrating the proposed controller with the fast frequency control scheme from \cite{ekomwenrenren2021hierarchical}, to dynamically demonstrate how our proposal can assist in providing fast ancillary services at the transmission level.

\appendix
\begin{pfof}{Theorem \ref{Thm:Stability}}
After eliminating the power set-points using \eqref{Eq:ParentSetPoints}, the dual update laws in Step 3 of Algorithm \ref{Alg:1} for the $i$th \ac{CA} can be compactly expressed as
\begin{equation}\label{Eq:CompactUpdate}
\mathbf{d}^+_i = \mc{P}_{\geq 0}\left(\mathbf{d}_i + \bs{\alpha}_i\left(\mathbf{C}_i\mathbf{y}_i + \mathbf{b}_i + \mathbf{D}_i\mathbf{x}_{\mathsf{P}(i)} - \mathbf{R}^{\rm d}_i \mathbf{d}_i\right)\right)
\end{equation}
where $\mathbf{C}_i,\mathbf{D}_i,\mathbf{b}_i$ are as in \eqref{Eq:BCD} and $\boldsymbol{\alpha}_i$, $\mathbf{R}^{\rm d}_i$ are as in Section \ref{Sec:Controller}. With $\mathbf{d} = (\mathbf{d}_1,\ldots,\mathbf{d}_N)$ the stacked vector of all dual variables for all \acp{CA}, we can express all updates together in vector form as
\begin{equation}\label{Eq:BigDual}
\mathbf{d}^+ = \mathcal{P}_{\geq 0}\left(
\mathbf{d} + \boldsymbol{\alpha}(\mathbf{C}\mathbf{y} + \mathbf{D}\boldsymbol{\mathcal{A}}^{\T}\mathbf{x} - \mathbf{R}^{\rm d}\mathbf{d})
\right)
\end{equation}
 Proceeding similarly, and using \eqref{Eq:LocalSensMatrices}, the DER set-point update \eqref{Eq:ArgminSimplified} can be compactly written as
\begin{equation}\label{Eq:CompactPrimal}
\mathbf{x}_i^{+} = \argmin_{\mathbf{x}_i \in \bs{\mathcal{X}}_i} f_i(\mathbf{x}_i) + \tfrac{r^{\rm p}_{i}}{2}\|\mathbf{x}_i\|_2^2 + (\mathbf{d}_i^+)^{\T}\mathbf{C}_i\mathbf{K}_i\mathbf{x}_i.
\end{equation}
The update \eqref{Eq:CompactPrimal} can be equivalently expressed as \cite{0157_Xingyu2018}
\begin{equation}\label{Eq:PrimalUpdateConvexConj}
\mathbf{x}_i^+ = \nabla \mathbf{F}_i^*(-\mathbf{K}_i^{\T}\mathbf{C}_i^{\T}\mathbf{d}_i^+),
\end{equation}
where $\mathbf{F}_i(\xi_i) = f_i(\xi) + \tfrac{r^{\rm p}_{i}}{2}\|\xi_i\|_2^2 + \mathbb{I}_{\bs{\mathcal{X}}_i}(\xi_i)$, with $\mathbb{I}_{\bs{\mathcal{X}}_i}(\xi_i)$ being the indicator function of the constraint set $\bs{\mathcal{X}}_i$ and $*$ denoting convex conjugation. By Assumptions in Sections \ref{Sec:SingleControlArea}, \ref{sec: LC Optimization Problem} \& \ref{Sec:Theory}, $\mathbf{F}_i$ is $(m_i+r^{\rm p}_{i})$-strongly convex, and hence $\mathbf{F}_{i}^*$ is continuously differentiable and $\frac{1}{m_i+r^{\rm p}_{i}}$-strongly smooth \cite{0157_Xingyu2018}. With $\mathbf{K}_{\rm d} = \mathrm{blkdiag}(\mathbf{K}_1,\ldots,\mathbf{K}_N)$, the stacked vector of all set-point updates for all areas can then be written as
\begin{equation}\label{Eq:BigPrimal}
\mathbf{x}^+ = \nabla \mathbf{F}^*(-\mathbf{K}_{\rm d}^{\T}\mathbf{C}^{\T}\mathbf{d}^+)
\end{equation}
where $\mathbf{F}(\boldsymbol{\xi}) = \sum_{i=1}^{N}\mathbf{F}_i(\xi_i)$. The closed-loop system is now described by the controller \eqref{Eq:BigDual},\eqref{Eq:BigPrimal} with the grid model \eqref{Eq:FullGrid}. Eliminating $\mathbf{y}$ and $\mathbf{x}$, we obtain the simplified representation
\begin{subequations}\label{Eq:BigDual2}
\begin{align}
\mathbf{d}^+ &= \mathcal{P}_{\geq 0}\left(
\mathbf{d} - \boldsymbol{\alpha}\boldsymbol{G}(\mathbf{d}) + \boldsymbol{\alpha}\mathbf{C}\mathbf{k}
\right)\\
\mathbf{G}(\mathbf{d}) &\define \mathbf{R}^{\rm d}\mathbf{d} - (\mathbf{C}\mathbf{K}+\mathbf{D}\boldsymbol{\mathcal{A}}^{\T})\nabla \mathbf{F}^*(-\mathbf{K}_{\rm d}^{\T}\mathbf{C}^{\T}\mathbf{d}).
\end{align}
\end{subequations}
Let $\mathbf{E}_{ii} = \mathbf{K}_{ii} - \mathbf{K}_i$. Then we may write
\[
\mathbf{K} = \mathbf{K}_{\rm d} + \begin{bmatrix}
\mathbf{E}_{11} & \cdots & \mathbf{K}_{1N}\\
\vdots & \ddots & \vdots\\
\mathbf{K}_{N1} & \cdots & \mathbf{E}_{NN}
\end{bmatrix} \define \mathbf{K}_{\rm d} + \boldsymbol{\Delta}.
\]
Now we write $\mathbf{G}(\mathbf{d}) = \sum_{k=1}^{4}\mathbf{G}_k(\mathbf{d})$, where $\mathbf{G}_1(\mathbf{d}) = \mathbf{R}^{\rm d}\mathbf{d}$ and
\[
\begin{aligned}
\mathbf{G}_2(\mathbf{d}) &= -\mathbf{C}\mathbf{K}_{\rm d}\nabla \mathbf{F}^*(-\mathbf{K}_{\rm d}^{\T}\mathbf{C}^{\T}\mathbf{d})\\
\mathbf{G}_3(\mathbf{d}) &= -\mathbf{C}\boldsymbol{\Delta}\nabla \mathbf{F}^*(-\mathbf{K}_{\rm d}^{\T}\mathbf{C}^{\T}\mathbf{d})\\
\mathbf{G}_4(\mathbf{d}) &= 
-\mathbf{D}\boldsymbol{\mathcal{A}}^{\T}\nabla \mathbf{F}^*(-\mathbf{K}_{\rm d}^{\T}\mathbf{C}^{\T}\mathbf{d})
\end{aligned}
\]
Let $\mathbf{d}, \mathbf{d}^{\prime}$ be two dual vectors, and let $\delta_i(\mathbf{d},\mathbf{d}^{\prime}) = (\mathbf{d}-\mathbf{d}^{\prime})^{\T}(\mathbf{G}_{i}(\mathbf{d}) - \mathbf{G}_{i}(\mathbf{d}^{\prime}))$. Since $\mathbf{R}^{\rm d} \succ 0$ and $\nabla \mathbf{F}^*$ is monotone \cite{0157_Xingyu2018}, we immediately have the bounds
\[
\begin{aligned}
\delta_1(\mathbf{d},\mathbf{d}^{\prime}) \geq \sum_{i=1}^{N} \|\mathbf{R}^{\rm d}_{i}\|_2 \|\mathbf{d}_i - \mathbf{d}_i^{\prime}\|_2^2, \quad \delta_2(\mathbf{d},\mathbf{d}^{\prime}) \geq 0.
\end{aligned}
\]
For $\mathbf{G}_3$ and $\mathbf{G}_4$, using strong smoothness of $\mathbf{F}_i$ one quickly obtains the bounds
\[
\begin{aligned}
|\delta_3(\mathbf{d},\mathbf{d}^{\prime})| &\leq \sum_{i,j=1}^{N}\nolimits \mathbf{Q}_{ij}\|\mathbf{d}_i-\mathbf{d}_i^{\prime}\|_2 \|\mathbf{d}_j-\mathbf{d}_j^{\prime}\|_2\\
|\delta_4(\mathbf{d},\mathbf{d}^{\prime})| &\leq \sum_{i,j=1}^{N}\nolimits \tilde{\mathbf{Q}}_{ij}\|\mathbf{d}_i-\mathbf{d}_i^{\prime}\|_2 \|\mathbf{d}_j-\mathbf{d}_j^{\prime}\|_2\\
\end{aligned}
\]
where $\mathbf{Q}, \tilde{\mathbf{Q}} \in \real^{N \times N}_{\geq 0}$ are defined element-wise as
\[
\begin{aligned}
\tilde{\mathbf{Q}}_{ij} &= \begin{cases}
\frac{\|\mathbf{C}_j\|_2\|\mathbf{D}_i\|_2 \|\mathbf{K}_j\|_2}{m_j + r^{\rm p}_{j}}, &(i,j) \in \mathcal{E}_{\rm CA}\\
0 &(i,j) \notin \mathcal{E}_{\rm CA}\\
\end{cases}\\
\mathbf{Q}_{ij} &= \begin{cases}
\frac{\|\mathbf{C}_i\|_2^2 \|\mathbf{K}_{ii} - \mathbf{K}_i\|_{2} \|\mathbf{K}_i\|_2}{m_i + r^{\rm p}_{i}}, &i = j\\
\frac{\|\mathbf{C}_i\|_2\|\mathbf{C}_j\|_2 \|\mathbf{K}_{ij}\|_2\|\mathbf{K}_j\|_2}{m_j + r^{\rm p}_{j}}, &i \neq j.
\end{cases}
\end{aligned}
\]
Putting things together, we obtain the lower bound
\[
\delta(\mathbf{d},\mathbf{d}^{\prime}) \geq \sum_{i,j=1}^{N}\nolimits\mathsf{M}_{ij}\|\mathbf{d}_i-\mathbf{d}_i^{\prime}\|_2 \|\mathbf{d}_j-\mathbf{d}_j^{\prime}\|_2
\]
where $\mathsf{M}$ is as defined in the theorem statement. Since $\mathsf{M} + \mathsf{M}^{\T} \succ 0$, it therefore holds that
\[
\delta(\mathbf{d},\mathbf{d}^{\prime}) \geq \tfrac{1}{2}\lambda_{\rm min}(\mathsf{M} + \mathsf{M}^{\T}) \|\mathbf{d}-\mathbf{d}^{\prime}\|_2^2,
\]
so we conclude that $\mathbf{G}$ is strongly monotone. One quickly observes that $\mathbf{G}$ is also Lipschitz continuous, with Lipschitz constant bounded by $\mathsf{L}$ as given in the theorem statement. By Theorem 12.1.2 in \cite{0153_facchinei2003finiteBookVol2}, it now follows that \eqref{Eq:BigDual2} possess a unique equilibrium point $\mathbf{d}^{\star} \geq 0$ and the equilibrium is globally exponentially stable if $\lambda_{\rm max}(\boldsymbol{\alpha})^2/\lambda_{\rm min}(\boldsymbol{\alpha}) < \lambda_{\rm min}(\mathsf{M} + \mathsf{M}^{\T})/\mathsf{L}^2$, which completes the proof. 
\end{pfof}

\normalem

\bibliographystyle{IEEEtran}
\bibliography{brevalias, refs, JWSP, Main,EE}


 





\end{document}

%% file: paperdefs.tex
\usepackage[plainpages=false]{hyperref}
\hypersetup{
    unicode=true,          
    colorlinks=true,       
    linkcolor=blue,        
    citecolor=blue,        
    breaklinks=true,
    pdftitle={Robust Feedback Systems and Integral Quadratic Constraints},
    pdfauthor={John W. Simpson-Porco},
    pdfsubject={},
    pdfkeywords={},
    baseurl={https://ece.uwaterloo.ca/~jwsimpso/multivariable.html}
}

\usepackage{etex}
\usepackage[vlined,ruled]{algorithm2e}
\usepackage[dvipsnames]{xcolor}
\usepackage{pgfpages}
\usepackage[cmex10]{amsmath}		
\usepackage{amssymb, mathrsfs}
\usepackage{cite}
\usepackage{url}

\usepackage{dsfont}
\usepackage{siunitx}
\usepackage{verbatim}									
\usepackage{mathtools}									
\usepackage{siunitx}									
\usepackage{mdframed}

\usepackage{graphicx}
\usepackage{epstopdf}

\usepackage[font=small]{caption}
\usepackage{subcaption}
\usepackage[activate={true,nocompatibility},final,tracking=true,kerning=true,spacing=true,factor=1100,stretch=10,shrink=10]{microtype}

\usepackage{array}
\usepackage{multirow}
\usepackage{enumerate}
\usepackage{booktabs}

\usepackage{algpseudocode}

\usepackage{soul}
\usepackage[printonlyused]{acronym}

\graphicspath{{./fig/}}

\usepackage{pgfplots}
\pgfplotsset{width=8cm,compat=newest}
\newcommand*{\damping}{0.006}%
\newcommand*{\freq}{25}%
\pgfmathsetmacro{\freqd}{sqrt(1-(\damping)^2)*\freq}%

\pgfplotsset{
    standard/.style={
    axis x line=middle,
    axis y line=middle,
    enlarge x limits=0.15,
	enlarge y limits=0.15,
	every axis plot post/.style={mark options={fill=black}},
	}
}

\pgfplotsset{%
    ,compat=1.12
    ,every axis x label/.style={at={(current axis.right of origin)},anchor=north west}
    ,every axis y label/.style={at={(current axis.above origin)},anchor=north east}
    }

\usepackage{tikz}
\usetikzlibrary{arrows}
\usetikzlibrary{decorations.pathmorphing}
\usetikzlibrary{decorations.markings}
\usetikzlibrary{snakes}
\usetikzlibrary{matrix}
\usetikzlibrary{calc}
\usetikzlibrary{patterns}
\usetikzlibrary{positioning}
\usetikzlibrary{shapes}
\usetikzlibrary{shapes.symbols}
\usetikzlibrary{shapes.geometric}
\usetikzlibrary{shadows.blur}
\usetikzlibrary{shapes.arrows}
\usetikzlibrary{shapes.multipart}
\usetikzlibrary{shapes.callouts}
\usetikzlibrary{shapes.misc}

\tikzstyle{every node}=[font=\small]
\tikzstyle{every path}=[line width=0.8pt,line cap=round,line join=round]

\usepackage[american,smartlabels]{circuitikz}
\ctikzset{bipoles/thickness=1}
\ctikzset{bipoles/length=0.8cm}
\ctikzset{bipoles/diode/height=.375}
\ctikzset{bipoles/diode/width=.3}
\ctikzset{tripoles/thyristor/height=.8}
\ctikzset{tripoles/thyristor/width=1}
\ctikzset{bipoles/vsourceam/height/.initial=.7}
\ctikzset{bipoles/vsourceam/width/.initial=.7}

\newcommand{\real}{\mathbb{R}}

\DeclareMathOperator*{\minimize}{minimize} 									

\DeclareMathOperator*{\argmin}{arg\,min}

\newcommand{\vect}[1]{\mathbbold{#1}}
\newcommand{\vones}[1][]{\vect{1}_{#1}}

\DeclareSymbolFont{bbold}{U}{bbold}{m}{n}
\DeclareSymbolFontAlphabet{\mathbbold}{bbold}

\newcommand{\T}{\mathsf{T}} 

\newcommand{\mc}[1]{\mathcal{#1}}
\newcommand{\mbf}[1]{\mathbf{#1}}
\newcommand{\bs}[1]{\boldsymbol{#1}}

\usepackage{tabularray}
\usepackage{ulem}

\newcommand\oprocendsymbol{\hbox{$\square$}}
\newcommand\oprocend{\relax\ifmmode\else\unskip\hfill\fi\oprocendsymbol}

\newtheorem{theorem}{Theorem}[section]

\newenvironment{pfof}[1]{\vspace{1ex}\noindent{\itshape Proof of
    #1:}\hspace{0.5em}} {\hfill\oprocend\vspace{1ex}}

\makeatletter
\newcommand{\vast}{\bBigg@{4}}
\newcommand{\Vast}{\bBigg@{5}}

\newcommand{\define}{\ensuremath{\triangleq}}



%% file: Main.bbl
\begin{thebibliography}{10}
\providecommand{\url}[1]{#1}
\csname url@samestyle\endcsname
\providecommand{\newblock}{\relax}
\providecommand{\bibinfo}[2]{#2}
\providecommand{\BIBentrySTDinterwordspacing}{\spaceskip=0pt\relax}
\providecommand{\BIBentryALTinterwordstretchfactor}{4}
\providecommand{\BIBentryALTinterwordspacing}{\spaceskip=\fontdimen2\font plus
\BIBentryALTinterwordstretchfactor\fontdimen3\font minus \fontdimen4\font\relax}
\providecommand{\BIBforeignlanguage}[2]{{%
\expandafter\ifx\csname l@#1\endcsname\relax
\typeout{** WARNING: IEEEtran.bst: No hyphenation pattern has been}%
\typeout{** loaded for the language `#1'. Using the pattern for}%
\typeout{** the default language instead.}%
\else
\language=\csname l@#1\endcsname
\fi
#2}}
\providecommand{\BIBdecl}{\relax}
\BIBdecl

\bibitem{abdul2012enhanced}
K.~H. Abdul-Rahman, H.~Alarian, M.~Rothleder, P.~Ristanovic, B.~Vesovic, and B.~Lu, ``Enhanced system reliability using flexible ramp constraint in {CAISO} market,'' in \emph{Proc. IEEE PESGM}, 2012, pp. 1--6.

\bibitem{FERC2020}
\BIBentryALTinterwordspacing
{Federal Energy Regulatory Commission}, ``Participation of distributed energy resource aggregations in markets operated by regional transmission organizations and independent system operators,'' Department of Energy, FERC Document 172 FERC ¶ 61,247, September 2020, docket No. RM18-9-000; Order No. 2222. [Online]. Available: \url{https://www.ferc.gov/sites/default/files/2020-09/E-1_0.pdf}
\BIBentrySTDinterwordspacing

\bibitem{dubey2023distribution}
A.~Dubey, S.~Paudyal \emph{et~al.}, ``Distribution system optimization to manage distributed energy resources (ders) for grid services,'' \emph{Foundations and Trends{\textregistered} in Electric Energy Systems}, vol.~6, no. 3-4, pp. 120--264, 2023.

\bibitem{ekomwenrenren2021hierarchical}
E.~Ekomwenrenren, Z.~Tang, J.~W. Simpson-Porco, E.~Farantatos, M.~Patel, and H.~Hooshyar, ``Hierarchical coordinated fast frequency control using inverter-based resources,'' \emph{IEEE Trans. Power Syst.}, vol.~36, no.~6, pp. 4992--5005, 2021.

\bibitem{ZT-EE-JWSP-EF-MP-HH:20l}
Z.~Tang, E.~Ekomwenrenren, J.~W. Simpson-Porco, E.~Farantatos, M.~Patel, and H.~Hooshyar, ``Measurement-based fast coordinated voltage control for transmission grids,'' \emph{IEEE Trans. Power Syst.}, vol.~36, no.~4, pp. 3416--3429, 2021.

\bibitem{ekomwenrenren2022integrated}
E.~Ekomwenrenren, Z.~Tang, J.~W. Simpson-Porco, E.~Farantatos, M.~Patel, H.~Hooshyar, and A.~Haddadi, ``An integrated frequency-voltage controller for next-generation power systems,'' in \emph{IEEE PES Innov. Smart Grid Tech. Conf. Europe}, Novi Sad, Serbia, Nov. 2022, pp. 1--6.

\bibitem{RRJ-AD:21}
R.~R. Jha and A.~Dubey, ``Network-level optimization for unbalanced power distribution system: Approximation and relaxation,'' \emph{IEEE Trans. Power Syst.}, vol.~36, no.~5, pp. 4126--4139, 2021.

\bibitem{SK-CM-PA-GH:21}
S.~Karagiannopoulos, C.~Mylonas, P.~Aristidou, and G.~Hug, ``Active distribution grids providing voltage support: The swiss case,'' \emph{IEEE Trans. Smart Grid}, vol.~12, no.~1, pp. 268--278, 2021.

\bibitem{HTN-DHC:23}
H.~T. Nguyen and D.-H. Choi, ``Decentralized distributionally robust coordination between distribution system and charging station operators in unbalanced distribution systems,'' \emph{IEEE Trans. Smart Grid}, vol.~14, no.~3, pp. 2164--2177, 2023.

\bibitem{AB-ED:19}
A.~{Bernstein} and E.~{Dall'Anese}, ``Real-time feedback-based optimization of distribution grids: A unified approach,'' \emph{IEEE Trans. Control Net. Syst.}, vol.~6, no.~3, pp. 1197--1209, 2019.

\bibitem{LB-MRJ-LS:23}
L.~Ballotta, M.~R. Jovanović, and L.~Schenato, ``Can decentralized control outperform centralized? the role of communication latency,'' \emph{IEEE Transactions on Control of Network Systems}, vol.~10, no.~3, pp. 1629--1640, 2023.

\bibitem{RG-FS-MP:21}
R.~Gupta, F.~Sossan, and M.~Paolone, ``Grid-aware distributed model predictive control of heterogeneous resources in a distribution network: Theory and experimental validation,'' \emph{IEEE Transactions on Energy Conversion}, vol.~36, no.~2, pp. 1392--1402, 2021.

\bibitem{LW-AD-AHG-AKS-NS:22}
L.~Wang, A.~Dubey, A.~H. Gebremedhin, A.~K. Srivastava, and N.~Schulz, ``Mpc-based decentralized voltage control in power distribution systems with ev and pv coordination,'' \emph{IEEE Trans. Smart Grid}, vol.~13, no.~4, pp. 2908--2919, 2022.

\bibitem{XC-ED-CZ-NL:20}
X.~Chen, E.~Dall’Anese, C.~Zhao, and N.~Li, ``Aggregate power flexibility in unbalanced distribution systems,'' \emph{IEEE Trans. Smart Grid}, vol.~11, no.~1, pp. 258--269, 2020.

\bibitem{XC-NL:21}
X.~Chen and N.~Li, ``Leveraging two-stage adaptive robust optimization for power flexibility aggregation,'' \emph{IEEE Transactions on Smart Grid}, vol.~12, no.~5, pp. 3954--3965, 2021.

\bibitem{AI-SK:23}
A.~Ingalalli and S.~Kamalasadan, ``Decentralized state estimation-based optimal integral model predictive control of voltage and frequency in the distribution system microgrids,'' \emph{IEEE Trans. Smart Grid}, vol.~14, no.~3, pp. 1790--1803, 2023.

\bibitem{BF-HI-CB:21}
B.~Fritz, H.~Ipach, and C.~Becker, ``Hierarchical online-optimization for the control of flexible loads and generators in distribution grids,'' in \emph{PESS 2021; Power and Energy Student Summit}, 2021, pp. 1--6.

\bibitem{GC-AB-RC-SZ:22}
G.~Cavraro, A.~Bernstein, R.~Carli, and S.~Zampieri, ``Feedback power cost optimization in power distribution networks with prosumers,'' \emph{IEEE Trans. Control Net. Syst.}, vol.~9, no.~4, pp. 1633--1644, 2022.

\bibitem{SF-GH-XZ-MC:21}
S.~Fan, G.~He, X.~Zhou, and M.~Cui, ``Online optimization for networked distributed energy resources with time-coupling constraints,'' \emph{IEEE Transactions on Smart Grid}, vol.~12, no.~1, pp. 251--267, 2021.

\bibitem{SN-YCC-LW:20}
S.~Nowak, Y.~C. Chen, and L.~Wang, ``Measurement-based optimal der dispatch with a recursively estimated sensitivity model,'' \emph{IEEE Transactions on Power Systems}, vol.~35, no.~6, pp. 4792--4802, 2020.

\bibitem{0155_Colombino}
M.~Colombino, J.~W. Simpson-Porco, and A.~Bernstein, ``Towards robustness guarantees for feedback-based optimization,'' in \emph{Proc. IEEE CDC}, 2019, pp. 6207--6214.

\bibitem{colombino2019online}
M.~Colombino, E.~Dall’Anese, and A.~Bernstein, ``Online optimization as a feedback controller: Stability and tracking,'' \emph{IEEE Trans. Control Net. Syst.}, vol.~7, no.~1, pp. 422--432, 2019.

\bibitem{LSPL-JWSP-EM:18l}
L.~S.~P. Lawrence, J.~W. Simpson-Porco, and E.~Mallada, ``Linear-convex optimal steady-state control,'' \emph{IEEE Trans. Autom. Control}, vol.~66, no.~11, pp. 5377--5384, Nov. 2021.

\bibitem{0154_ORTMANN2020106782}
L.~Ortmann, A.~Hauswirth, I.~Caduff, F.~Dörfler, and S.~Bolognani, ``Experimental validation of feedback optimization in power distribution grids,'' \emph{Electric Power Syst. Research}, vol. 189, p. 106782, 2020.

\bibitem{EDA-SVD-GBG:15}
E.~Dall'Anese, S.~V. Dhople, and G.~B. Giannakis, ``Regulation of dynamical systems to optimal solutions of semidefinite programs: Algorithms and applications to ac optimal power flow,'' in \emph{Proc. {ACC}}, Chicago, IL, USA, July 2015, pp. 2087--2092.

\bibitem{GB-JC-JIP-ED:21}
G.~Bianchin, J.~Cort\'{e}s, J.~I. Poveda, and E.~Dall'Anese, ``Time-varying optimization of {LTI} systems via projected primal-dual gradient flows,'' \emph{IEEE Trans. Control Net. Syst.}, vol.~9, no.~1, pp. 474--486, 2022.

\bibitem{AB-JC-YC-JW:23}
A.~Bernstein, J.~Comden, Y.~Chen, and J.~Wang, ``Time-varying feedback optimization for quadratic programs with heterogeneous gradient step sizes,'' 2023.

\bibitem{0165_Han2014}
X.~Han, A.~M. Kosek, D.~E. Morales~Bondy, H.~W. Bindner, S.~You, D.~V. Tackie, J.~Mehmedalic, and F.~Thordarson, ``Assessment of distribution grid voltage control strategies in view of deployment,'' in \emph{2014 IEEE International Workshop on Intelligent Energy Systems (IWIES)}, 2014, pp. 46--51.

\bibitem{0166_Utkarsh2022}
K.~Utkarsh, F.~Ding, X.~Jin, M.~Blonsky, H.~Padullaparti, and S.~P. Balamurugan, ``A network-aware distributed energy resource aggregation framework for flexible, cost-optimal, and resilient operation,'' \emph{IEEE Trans. Smart Grid}, vol.~13, no.~2, pp. 1213--1224, 2022.

\bibitem{BS-EPRI:18}
\BIBentryALTinterwordspacing
B.~Seal, ``Understanding \text{DERMS},'' \emph{Electric Power Research Institute (EPRI) Technical Update (3002013049)}, vol. 2018, 2018. [Online]. Available: \url{https://www.epri.com/research/products/000000003002013049}
\BIBentrySTDinterwordspacing

\bibitem{BD-EPRI:13}
\BIBentryALTinterwordspacing
B.~Deaver, ``Distribution management systems planning guide,'' \emph{Electric Power Research Institute (EPRI), Technical Update (1024385)}, 2013. [Online]. Available: \url{https://www.epri.com/research/products/1024385}
\BIBentrySTDinterwordspacing

\bibitem{AM-TH-JW-RS-NK-XL-JR-AP-SV:17}
\BIBentryALTinterwordspacing
A.~Pratt, S.~Veda, A.~Maitra, T.~Hubert, J.~Reilly, J.~Wang, R.~Singh, N.~Kang, and X.~Lu, ``\text{DMS} advanced applications for accommodating high penetrations of \text{DERs} and microgrids,'' National Renewable Energy Lab.(NREL), Golden, CO (United States), Tech. Rep., 2017. [Online]. Available: \url{https://www.nrel.gov/docs/fy17osti/67775.pdf}
\BIBentrySTDinterwordspacing

\bibitem{0057_FPL_Bernstein_followup}
A.~Bernstein and E.~Dall'Anese, ``Linear power-flow models in multiphase distribution networks,'' in \emph{IEEE PES Innov. Smart Grid Tech. Conf. Europe}, 2017, pp. 1--6.

\bibitem{0059_FPL_Bernstein_Parent}
A.~Bernstein, C.~Wang, E.~Dall’Anese, J.-Y. Le~Boudec, and C.~Zhao, ``Load flow in multiphase distribution networks: Existence, uniqueness, non-singularity and linear models,'' \emph{IEEE Trans. Power Syst.}, vol.~33, no.~6, pp. 5832--5843, 2018.

\bibitem{FarhatSupplOnlineMat}
\BIBentryALTinterwordspacing
I.~Farhat, ``Supplementary online materials for ``transmission and distribution networks coordination: A novel multi-layered control structure'' manuscript.'' [Online]. Available: \url{https://scholar.google.com/citations?user=fMXYx_MAAAAJ}
\BIBentrySTDinterwordspacing

\bibitem{0156_Sumin2022}
V.~I. Sumin and M.~I. Sumin, ``On the iterative regularization of the lagrange principle in convex optimal control problems for distributed systems of the volterra type with operator constraints,'' \emph{Differential Equations}, vol.~58, no.~6, pp. 791--809, Jun 2022.

\bibitem{0158_Agarwal2022}
A.~Agarwal, J.~W. Simpson-Porco, and L.~Pavel, ``Game-theoretic feedback-based optimization,'' \emph{IFAC-PapersOnLine}, vol.~55, no.~13, pp. 174--179, 2022, iFAC NecSys Workshop.

\bibitem{0159_PavelBook2012}
L.~Pavel, \emph{Game theory for control of optical networks}.\hskip 1em plus 0.5em minus 0.4em\relax Springer Science \& Business Media, 2012.

\bibitem{GB-DLM-MHDB-SB-JL-FD:21}
G.~Belgioioso, D.~Liao-McPherson, M.~H. de~Badyn, S.~Bolognani, J.~Lygeros, and F.~D\"{o}rfler, ``Sampled-data online feedback equilibrium seeking: Stability and tracking,'' in \emph{Proc. IEEE CDC}, Austin, TX, USA, 2021, pp. 2702--2708.

\bibitem{0160_DILEEP2020}
G.~Dileep, ``A survey on smart grid technologies and applications,'' \emph{Renewable Energy}, vol. 146, pp. 2589--2625, 2020.

\bibitem{0161_ALI2017}
I.~Ali and S.~S. Hussain, ``Control and management of distribution system with integrated ders via iec 61850 based communication,'' \emph{Engineering Science and Technology, an International Journal}, vol.~20, no.~3, pp. 956--964, 2017.

\bibitem{Farhat_MATDSSApplication}
\BIBentryALTinterwordspacing
I.~Farhat, ``{MATDSS} application v0.87.'' [Online]. Available: \url{https://scholar.google.com/citations?user=fMXYx_MAAAAJ}
\BIBentrySTDinterwordspacing

\bibitem{KJA-TH:06}
K.~{\AA}str{\"o}m and T.~H{\"a}gglund, \emph{Advanced PID Control}.\hskip 1em plus 0.5em minus 0.4em\relax ISA-The Instrumentation, Systems, and Automation Society; Research Triangle Park, NC 27709, 2006.

\bibitem{VH-LH-XH-EPA-FD:23}
V.~Häberle, L.~Huang, X.~He, E.~Prieto-Araujo, and F.~Dörfler, ``Dynamic ancillary services: From grid codes to transfer function-based converter control,'' 2023.

\bibitem{JC-Jw-AB:23}
J.~Comden, J.~Wang, and A.~Bernstein, ``Adaptive primal-dual control for distributed energy resource management,'' 2023.

\bibitem{0157_Xingyu2018}
X.~Zhou, ``On the {F}enchel duality between strong convexity and lipschitz continuous gradient,'' 2018.

\bibitem{0153_facchinei2003finiteBookVol2}
F.~Facchinei and J.-S. Pang, \emph{Finite-dimensional variational inequalities and complementarity problems}.\hskip 1em plus 0.5em minus 0.4em\relax Springer, 2003, vol.~2.

\end{thebibliography}
